\documentclass[journal,twocolumn,10pt,twoside]{IEEEtranTCOM}
\usepackage{bbm}
\normalsize
\usepackage{cite}
\ifCLASSINFOpdf
\usepackage[pdftex]{graphicx}
\graphicspath{{Figs/}}
\else
\usepackage[dvips]{graphicx}
\graphicspath{{Figs/}}
\fi

\usepackage{amsmath}
\usepackage{amssymb}
\usepackage{algorithmic}
\usepackage{algorithm}
\usepackage{array}
\usepackage{mathtools}
\usepackage{amsthm}

\newcommand{\mymin}{\operatornamewithlimits{\min}}
\DeclareMathOperator{\sign}{sign}
\newtheorem{assum}{Assumption}
\newtheorem{theo}{Theorem}

\usepackage{blkarray}
\usepackage{balance}
\usepackage{mathrsfs}
\usepackage[super]{nth}
\DeclareMathOperator\erfc{erfc}

\usepackage{color,soul} 
\definecolor{red}{rgb}{1,0,0} 

\definecolor{lightblue}{rgb}{.90,.95,1} 
\sethlcolor{yellow} 

\begin{document}
	
	\title{Performance Bounds and Estimates\\ for Quantized LDPC Decoders}
	
	\author{Homayoon~Hatami,~\IEEEmembership{Member,~IEEE,}
		David~G.~M.~Mitchell,~\IEEEmembership{Senior Member,~IEEE,}\\
		Daniel~J.~Costello,~Jr.,~\IEEEmembership{Life Fellow,~IEEE,}
		and~Thomas~E.~Fuja~\IEEEmembership{Fellow,~IEEE}
		\thanks{This work was supported in part by the National Science Foundation under Grant Nos. ECSS-1710920 and OIA-1757207. Some of the material was previously presented at the IEEE International
			Symposium on Information Theory, Barcelona, Spain, July 2016 and at
			the Allerton Conference on Communication, Control, and Computing, Monticello, IL, USA, October 2017.}%
		\thanks{Homayoon~Hatami, Daniel~J.~Costello,~Jr, and Thomas~E.~Fuja are with the Department
			of Electrical  Engineering, University of Notre Dame, Notre Dame,
			IN, 46545 USA (e-mail: hhatami@nd.edu;~~~costello.2@nd.edu;~~~tfuja@nd.edu).}
		\thanks{David~G.~M.~Mitchell is with the Klipsch School of Electrical and Computer Engineering, New Mexico State University, Las Cruces, NM, 88003 USA (email: dgmm@nmsu.edu).}}
	\markboth{IEEE Transactions on Communications}%
	{Submitted paper}
	
	\maketitle
	\begin{abstract}
	The performance of low-density parity-check (LDPC) codes at high signal-to-noise ratios (SNRs) is known to be limited by the presence of certain sub-graphs that exist in the Tanner graph representation of the code, for example trapping sets and absorbing sets. This paper derives a lower bound on the frame error rate (FER) of any LDPC code containing a given problematic sub-graph, assuming a particular message passing decoder and decoder quantization. A crucial aspect of the lower bound is that it is code-independent, in the sense that it can be derived based only on a problematic sub-graph and then applied to any code containing it. Due to the complexity of evaluating the exact bound, assumptions are proposed to approximate it, from which we can estimate decoder performance. Simulated results obtained for both the quantized sum-product algorithm (SPA) and the quantized min-sum algorithm (MSA) are shown to be consistent with the approximate bound and the corresponding performance estimates. Different classes of LDPC codes, including both structured and randomly constructed codes, are used to demonstrate the robustness of the approach.

	\end{abstract}
	
	
	\IEEEpeerreviewmaketitle

\section{Introduction}

\IEEEPARstart{L}{ow}-density parity-check (LDPC) codes~\cite{gallager1} are a class of error correcting codes with asymptotic performance approaching the Shannon limit. However, practical LDPC decoders, such as those that implement message-passing algorithms based on belief propagation (BP), can introduce an \textit{error floor} that limits error probability at high {signal-to-noise} ratios (SNRs).  A number of structures in a code’s Tanner graph representation have been identified as significant factors in error floor performance – \textit{e.g.}, \textit{near-codewords}~\cite{MacKay03}, \textit{trapping sets}~\cite{richardson03}, and \textit{absorbing sets}~\cite{dolecek10}. Absorbing sets are known to be problematic in a variety of LDPC codes and stable under bit flipping decoding~\cite{tomasoni17,Kyung12}. Other classes of trapping sets, such as \textit{elementary trapping sets} and \textit{leafless elementary trapping sets}, have been shown to be the dominant cause of the error floor for certain codes~\cite{LETS,LETS1,LETS2,LETS3}.

Several papers have addressed the problem of predicting the error floor performance of LDPC codes on the {additive white Gaussian noise (AWGN)} channel based on the existence of these problematic structures. In~\cite{richardson03}, Richardson proposed a variation of importance sampling to estimate the frame error rate (FER) of a code based on trapping sets. In~\cite{dolecek09}, an error floor estimate was introduced based on the \textit{dominant} absorbing sets (those empirically determined to cause most errors) in structured array-based codes, and the results were compared to those derived from importance sampling. In~\cite{alerton17}, a method similar to~\cite{dolecek09} was applied to the min-sum algorithm (MSA). In~\cite{bani13}, the contribution of the shortest cycles in a code’s graph was used to estimate its performance. Also,~\cite{sun04} and~\cite{schelegel13} developed a state-space model for a code’s dominant absorbing sets to estimate its FER. Later,~\cite{butler14} applied this method to the case where the log-likelihood-ratios (LLRs) used for decoding are constrained to some maximum saturation value. Each of these references considered the problematic structures of a particular code. In contrast, the authors of~\cite{tomasoni17} derived a real-valued threshold associated with a particular absorbing set irrespective of the code; the threshold indicates if the absorbing set can be “deactivated” and hence not contribute to the FER at high SNR in any code that contains it.
	
This paper obtains \textit{sub-graph specific}, or \textit{code-independent}, lower bounds on the performance of an LDPC code when a finite precision (quantized) LDPC decoder is used. These bounds are \textit{general}, in that they apply to \textit{any code} containing a particular problematic sub-graph; however, calculating the bound is complex, so we introduce assumptions and approximations to simplify its calculation, resulting in what we call an \textit{approximate lower bound}. Given a description of a dominant problematic sub-graph and its multiplicity in a code, an estimate of the resulting FER performance is obtained. Extensive simulation results justify the validity of the assumptions and approximations used for various decoders, quantizers, problematic sub-graphs, and codes.

We first create a simplified model for the Tanner graph of a code containing a particular problematic sub-graph; this model captures the structure of the code outside the sub-graph with a single edge connected to each check node incident to a variable node inside the sub-graph. We use this model to identify the sets of quantized received channel LLR values observed at the sub-graph’s variable nodes that cannot be corrected even under the most favorable LLR conditions for the variable nodes outside the sub-graph. These sets are deterministic for a given sub-graph, \textit{i.e.}, they cause a decoding error regardless of the channel SNR, and thus they can be used to lower bound the FER performance of any code containing that sub-graph, and it is not necessary to re-derive the sets for every SNR. Furthermore, deriving these sets is typically much faster than performing a Monte-Carlo simulation for a particular SNR. The probabilities of these sets of received values are functions of the SNR and can be derived analytically; the same bound can be used for two different codes with the same absorbing set but different rates, one bound being a simple SNR-derived shift of the other.  We refer to these bounds as ``code-independent''. To verify the accuracy of the lower bound and the corresponding performance estimates, we have considered a variety of codes, including array-based codes of different rates, Euclidean Geometry codes, Tanner codes, and randomly constructed codes for both {sum-product algorithm (SPA)} and MSA decoders and uniform and non-uniform quantizers. Our focus is on absorbing sets, since they have been well-studied in the literature.

\section{Background}

\subsection{LDPC Codes/Quantized Decoders}\label{sec:mapping}	
Assume that a codeword  ${\mathbf{y}} = \left(y_1,y_2,\ldots,y_n\right)$ is binary phase shift keying (BPSK) modulated such that each zero is mapped to $+1$ and each one is mapped to $-1$. The modulated signal is transmitted over an AWGN channel with mean $0$ and standard deviation $\sigma$. The received samples  from the channel are multiplied by $2/\sigma^2$ to form the channel LLR vector $\mathbf{\tilde{r}}$ corresponding to ${\mathbf{y}}$. 
	As a result, for $i = 1,2,\ldots,n$, the element of $\mathbf{\tilde{r}}$ corresponding to $y_i$, denoted ${\tilde{r_i}}$, has a Gaussian distribution with mean  $2/\sigma^2$ or $-2/\sigma^2$, depending on whether {the modulated symbol is $+1$ or $-1$}, respectively. The standard deviation of each ${\tilde{r_i}}$ is $2/\sigma$, and since LDPC codes are linear, we can assume the transmission of  the all-zero codeword. Therefore, 
	\begin{equation}
	\tilde{r_i} \sim \mathcal{N}\left(\frac{2}{\sigma^2},\frac{4}{\sigma^2}\right),~~ i=1,2,\ldots,n,
	\label{eqn:1}
	\end{equation}
	where $\mathcal{N}(m,\sigma^2)$ is the Gaussian distribution with mean $m$ and standard deviation ${\sigma}$.
	
Let the sets $V = \{v_1,v_2, \ldots, v_n\}$ and  $C = \{c_1,c_2,\ldots,c_m\}$ represent the set of variable nodes and check nodes, respectively, of a bipartite Tanner graph representation $\mathcal{G}$ of an LDPC code parity-check matrix. 	In practical decoder implementations, the channel LLRs and variable node and check node LLRs must be quantized, and the calculations at check nodes and variable nodes are implemented with finite precision. At a given iteration, let $\mathbb{V}_{i\rightarrow j}$ represent the quantized LLR passed from $v_i$ to $c_j$. Similarly, let $\mathbb{C}_{j\rightarrow i}$ represent the quantized LLR passed from $c_j$ to $v_i$. The set of check nodes that are neighbors (connected) to $v_i$ are {denoted by} $N(v_i)$, and the set of variable nodes that are neighbors to  $c_j$ are {denoted by} $N(c_j)$. To initialize decoding, each variable node $v_i$ passes a quantized version of ${\tilde{r_i}}$, {denoted by} ${{r_i}}$, to the check nodes in $N(v_i)$. At the check nodes, the LLR passed from $c_j$ to $v_i$ is calculated as follows for quantized SPA and MSA decoders:

\begin{itemize}
	\item Quantized SPA:
	The check node operation  can be written as 
	\begin{equation}
	\begin{split}
	\mathbb{C}_{j\rightarrow i} 
	&  =  \prod_{i'\in N(c_j)\setminus i}\sign\left ( \mathbb{V}_{i'\rightarrow j} \right )  \\
	&\cdot \Phi_2\left ( \sum_{i'\in N(c_j)\setminus i} \Phi_1\left ( \left | \mathbb{V}_{i'\rightarrow j} \right | \right ) \right ),
	\end{split}
	\label{eqn:5}
	\end{equation}
	where the two functions $\Phi_1(x)$ and $\Phi_2(x)$ are defined as $	\Phi_1(x) = Q\left(\Phi(x)\right)$ and $\Phi_2(x) = Q\left(\Phi^{-1}\left(Q\left(x\right)\right)\right)$, $\Phi(x) = \log \left( \frac{e^{x}+1}{e^{x}-1} \right)$, and  the function $Q(x)$ returns the quantized value of $x$.
 In~\cite{dolecek09}, it is shown that this quantized  implementation suffers from a significant error floor, \textit{i.e.}, at high SNRs there is little additional reduction in the FER as the channel quality improves.
		
	\item Quantized MSA:
	The check node computation simplifies to
	\begin{equation}
	\mathbb{C}_{j\rightarrow i} =    \prod_{i'\in N(c_j)\setminus i}\sign\left ( \mathbb{V}_{i'\rightarrow j} \right )  \cdot \mymin_{i'\in N(c_j)\setminus i} \left | \mathbb{V}_{i'\rightarrow j} \right |.
	\label{eqn:88}
	\end{equation}
	The MSA is an approximation of the SPA that reduces implementation complexity. 
\end{itemize}

	For both the SPA and MSA, at the variable nodes, the hard decision estimate  ${\hat {\mathbf{y}}} = \left(\hat{y_1},\hat{y_2},\ldots,\hat{y_n}\right)$ is checked to see if it is a valid codeword, where  $\hat{y_i} = 0$ iff 
	\begin{equation}
		{{r_i}}+  \sum_{j'\in N(v_i)} \mathbb{C}_{j'\rightarrow i}>0.
		\label{eqn:3}
	\end{equation}
	If ${\hat {\mathbf{y}}}$  is not a valid codeword and fewer than $p$ iterations have been carried out, the next iteration is performed and the LLRs passed from the variable nodes to the check nodes are 
	\begin{equation}
		\mathbb{V}_{i\rightarrow j} = Q\left({{r_i}}+  \sum_{j'\in N(v_i)\setminus j} \mathbb{C}_{j'\rightarrow i}\right).
		\label{eqn:4}
	\end{equation}

\subsection{Trapping Sets \& Absorbing Sets}	
\begin{figure}[t]
		\centering
		\includegraphics[width=0.7\columnwidth]{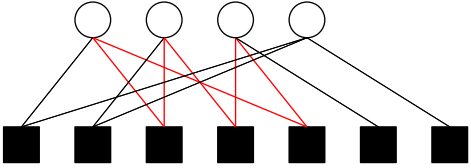}
		\caption{An illustration of a $\left(4,2\right)$ absorbing set with girth $6$. This sub-graph can also be referred to as an elementary trapping set or a leafless elementary trapping set.}\label{Aset4_2}
		
	\end{figure}

Let ${{A}} = \{v_1(A),v_2(A),\ldots,v_a(A)\}$ denote a subset of $V$ of cardinality $a$.   Let $A_\text{even}$ and $A_\text{odd}$ represent the subsets of check nodes connected to variable nodes in ${{{A}}}$ with even and odd degrees, respectively, where {$\left | A_\text{odd} \right | = b$}. {Let the sub-graph induced by ${A}$ be $\mathcal{G}(A) = (A\cup C(A) , E)$, where $C(A) =\left\{ c_1(A),c_2(A),\ldots,c_\gamma(A) \right\} $ represents the check nodes connected to ${A}$, $|C(A)| = \gamma$, and $E$ represents the set of edges connecting ${A}$ to $C(A)$.} The sub-graph of $\mathcal{G}$ that is induced by $A$ is called an $\left(a,b\right)$ \emph{trapping set}, with graphical representation $\mathcal{G}{(A)}$. ${{{A}}}$ is further defined to induce an $(a,b)$ \emph{absorbing set} if each variable node in ${{{A}}}$ is connected to fewer check nodes in $A_\text{odd}$ than $A_\text{even}$. As an illustration, Fig.~\ref{Aset4_2} shows a $(4,2)$ absorbing set with $b=2$ degree-one check nodes in $A_\text{odd}$, where each of the $a=4$ variable nodes in ${{{A}}}$ is connected to fewer elements in $A_\text{odd}$ than $A_\text{even}$.  This $(4,2)$ absorbing set is a structure that appears often in $(3,K)$-regular LDPC codes, for example, and we see that it contains a \emph{cycle} of length six (the highlighted edges in the figure).  The \emph{girth} of an absorbing set is the length of its shortest cycle, and it can be readily observed that the girth of the absorbing set in  Fig.~\ref{Aset4_2} is  six.
	
	Other classifications of problematic sub-graphs have been referred to as elementary trapping sets (ETS), which contain only degree-1 and degree-2 check nodes~\cite{LETS2}, and leafless elementary trapping sets (LETS), in which each variable node is connected to at least two even-degree check nodes~\cite{LETS}. As such, Fig.~\ref{Aset4_2} can also be referred to as a $(4,2)$ ETS or LETS.
		
\subsection{Quantizers}	
Since  quantized decoding may have different performance characteristics than unquantized decoding, considering the effect of quantization on decoder performance is of great importance:
\begin{itemize}  
	\item Uniform Quantization:
	Following convention, we let $\text{Q}_{{q_1}.{q_2}}$ denote a quantizer that represents each message with $q = q_1 + q_2 + 1$ bits:  $q_1$ bits to represent the integer part of the message, $q_2$ bits to represent the fractional part, and one bit to represent the sign. In this case, there are $t = 2^q$ quantization levels, where the levels (\textit{i.e.}, the quantized message values) range from ${\ell}_{1}={-2^{q_1}}$ to ${\ell}_{{t}}={2^{ {q_1}}}-{2^{ - {q_2}}}$, with step size $\Delta= {2^{ - {q_2}}}$ between levels. The quantizer thresholds are equidistant between the levels and range from $b_1$ to $b_{t-1}$, where ${b_i} = \frac{{{\ell}_i}+{{\ell}_{i+1}}}{2}$ for $i \in \{1,2,\ldots,t-1\}$.
	\item Quasi-Uniform Quantization:
	  In~\cite{siegel14}, the authors proposed a non-uniform quantizer, denoted as ``quasi-uniform'' due to its structure, which uses $q$ bits for uniform quantization, thus maintaining precision, plus an extra bit to increase the range of the quantizer compared to a $q+1$ bit uniform quantizer. It is shown in~\cite{siegel14} that the increased range of this quantizer improves the error-floor performance.
	\end{itemize}	
	
\section{System model }\label{sec:system}
In this section, we propose a general model for  representing a problematic sub-graph in an arbitrary code.  We also formulate expressions for the quantized LLR values received at the variable nodes and check nodes  in the sub-graph. As mentioned earlier, we focus on absorbing sets as our sub-graph of interest in the development of our system model; however, the system model can be generalized in a straightforward manner to any sub-graph. 

\subsection{Absorbing Set Model}
	
\begin{figure}[t]
		\centering
		\includegraphics[width=0.8\columnwidth]{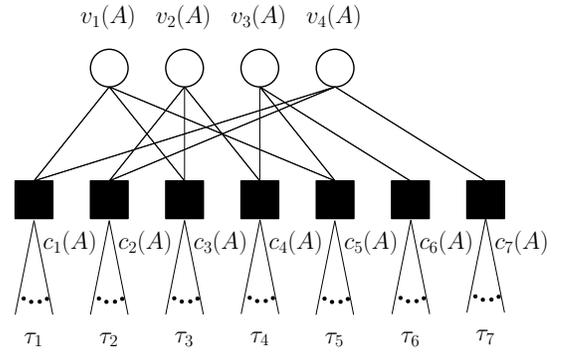}
		\caption{An illustration of a $(4,2)$ absorbing set with an unspecified number of edges  connected to each check node.}\label{Aset4_2_L}
	\end{figure}

We consider the general case of an $(a,b)$ absorbing set  with an unspecified number of edges connected to each of its check nodes. The variable nodes are represented by ${A}\subset V$.  We partition the edges connected to each $c_j(A)$ into two groups depending on whether they connect to a variable node in $A$ or $V\setminus A$.   We denote the neighboring nodes of $c_j(A)$  in $A$ as $N'(c_j(A))$ and the neighboring nodes of $c_j(A)$ in $V\setminus A$ as $N''(c_j(A))$. If there are {$\rho_j \geq 1$}  edges connected to $A$ and {$\tau_j \geq 0$} edges connected to $V\setminus A$, then $|N'(c_j(A))| = \rho_j$ and $|N''(c_j(A))| = \tau_j$.  In  Fig.~\ref{Aset4_2_L}, a $(4,2)$ absorbing set is illustrated in which $\rho_j = 2$, $j = 1,2,\ldots,5$, $\rho_6=\rho_7 = 1$, and $\tau_j$ is arbitrary for $j = 1,2,\ldots, 7$ {(note that $\tau_j$ can be zero)}.

\begin{figure}[t]
		\centering
		\includegraphics[width=0.8\columnwidth]{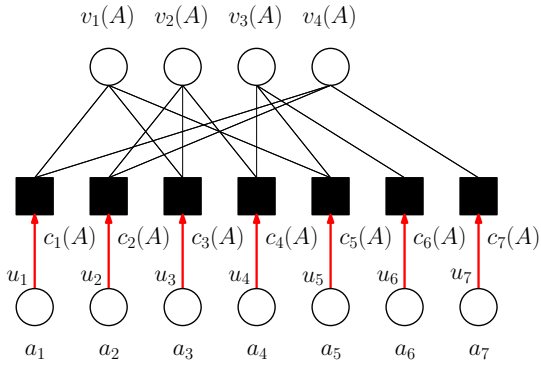}
		\caption{An illustration of a $(4,2)$ absorbing set decoder graph $\mathcal{D}(A)$ with single edges connected from auxiliary variable node $a_j$ to each check node $c_j(A)$, where each $u_j$ represents the LLR input to check node $c_j(A)$ from outside $\mathcal{G}(A)$.}\label{Aset4_2_e}
	\end{figure}

To simplify the calculation of the LLRs sent from each check node $c_j(A)$ to the variable nodes {$v_i(A) \in A$ in the case where $\tau_j > 0$,} we represent the $\tau_j$ edges  from the variable nodes in $N''(c_j(A))$ with a single edge (see Fig.~\ref{Aset4_2_e}). This edge has an LLR $u_j$ that is a function of all the external LLRs coming from the set $N''(c_j(A))$ to $c_j(A)$ and can be derived as follows: 
	\begin{itemize}
		\item SPA:
		\begin{equation}
		 \left\{\begin{matrix}
		\sign(u_j) = \prod_{i'\in N''(c_j)}\sign\left ( \mathbb{V}_{i'\rightarrow j} \right ), \\
		\Phi_1\left ( \left | u_j \right |    \right ) =   \sum_{i'\in N''(c_j)}\Phi_1\left ( \left | \mathbb{V}_{i'\rightarrow j} \right | \right ).
		\end{matrix}\right.
		\end{equation}\label{501}
		\item MSA: 
		\begin{equation}
		{u}_{j} \triangleq \prod_{i'\in N''(c_j(A))}\sign\left ( \mathbb{V}_{i'\rightarrow j} \right )  \cdot \mymin_{i'\in N''(c_j(A))} \left | \mathbb{V}_{i'\rightarrow j} \right |.
		\label{eqn:890}
		\end{equation}
	\end{itemize}
	 LLR  ${u}_j$ can then be used in equations~\eqref{eqn:5} and~\eqref{eqn:88}, in conjunction with the internal LLRs coming from the set $N'(c_j(A))$, to form the LLRs sent from $c_j(A)$ to the variable nodes in $A$. {(Note that if $\tau_j = 0$ for any $j$, then the single edge representation described above is not necessary since outgoing messages from $c_j(A)$ will be a function only of internal messages from $v_i(A) \in A$.)}
	 
This simplification, where we consider only one {external} edge connected to {check nodes $c_j(A)$ with $\tau_j >0$} in $\mathcal{G}(A)$ from outside  the absorbing set, is depicted in Fig.~\ref{Aset4_2_e} for a $(4,2)$ absorbing set. We refer to this graph as {the} \textit{absorbing set decoder graph} $\mathcal{D}(A) = (A\cup C(A) \cup A', E \cup E')$, where $A' = \{a_1, a_2,\ldots,a_{\kappa}\}$ is the set of \textit{auxiliary variable nodes}{, $\kappa \leq \gamma$ corresponds to the number of check nodes in $C(A)$ with $\tau_j>0$,} and $E'$ is the set of  single edges connecting each ${a}_j$ to $c_j(A)$. We also refer to a decoder operating on $\mathcal{D}(A)$ as an \emph{absorbing set decoder}.\footnote{{We remind the reader that the concept of an absorbing set decoder can be applied to any sub-graph of interest.}} Later, we will use an absorbing set decoder operating on $\mathcal{D}(A)$ to develop a lower bound on the FER of any code containing $\mathcal{G}(A)$. No detailed  information about the code containing the absorbing set is required in this approach, except the code rate, which is needed to determine the channel SNR in terms of ${E_\text{b}}/{N_\text{0}}$. In the next two sub-sections, we discuss how the possible inputs to the variable nodes and check nodes of a quantized absorbing set decoder are determined.

\subsection{Variable Node Inputs}
Let $\mathbf{r} = (r_1,r_2,\ldots,r_n)$ denote the quantized  version of $\tilde{\mathbf{r}}$ corresponding to the variable nodes in ${V}$ as described in~\eqref{eqn:1}. The portion of $\mathbf{r}$ (respectively $\tilde{\mathbf{r}}$) corresponding to the variable nodes of $A$ is {denoted by} $\mathbf{s} = (s_1,s_2,\ldots,s_a)$ (resp. $\tilde{\mathbf{s}} = (\tilde{s_1},\tilde{s_2},\ldots,\tilde{s_a})$).  Each element of $\mathbf{s}$, {denoted by} $s_m$, $m = 1,2,\ldots,a$, can take one of $t= 2^q$ values for a $q$-bit quantizer. These values are labeled ${\ell}_1$ to ${\ell}_t$, from smallest to largest. The quantizer boundaries are represented by $e_1$ to $e_{t-1}$. The probability that $s_m$ takes on the value ${\ell}_{k}$, $k = 1,2,\ldots,t$, is equal to the probability that $e_{k-1} <{\tilde{s}_m} < e_k$, where $e_0 = -\infty$ and $e_t = \infty $. For the AWGN channel, this probability is given by
	\begin{multline}
	\Pr(s_m = {\ell}_k) =\\
	\left\{\begin{matrix}
	\frac{1}{2}\erfc\left ( \frac{e_{t-1}-\frac{2}{{\sigma}^2}}{\frac{2\sqrt{2}}{{\sigma}}} \right ),& \text{if } k={t} \\
	\frac{1}{2}\erfc\left ( \frac{e_{k-1}-\frac{2}{{\sigma}^2}}{\frac{2\sqrt{2}}{{\sigma}}} \right )-
	\frac{1}{2}\erfc\left ( \frac{e_{k}-\frac{2}{{\sigma}^2}}{\frac{2\sqrt{2}}{{\sigma}}} \right), & \text{if }  1<k<t\\ 
	1-\frac{1}{2}\erfc\left ( \frac{e_{1}-\frac{2}{{\sigma}^2}}{\frac{2\sqrt{2}}{{\sigma}}} \right ),& \text{if }  k=1
	\end{matrix}\right..
	\label{eqn:7}
	\end{multline}   
\noindent where $\erfc(\cdot)$ represents the complementary error function of Gaussian statistics.
	The vector $\mathbf{s}$ can take on $t^a$ different values, representing the possible combinations of quantizer levels $\mathbf{x}_i = ({\ell}_{i_1}, {\ell}_{i_2},\ldots, {\ell}_{i_a})$, $i_1, i_2, \ldots, i_a \in \{1,2,\ldots,t\}$, $i = 1,2,\ldots, t^{ a}$. The set of all possible vectors $\mathbf{s}$ is {denoted by} $X$, and the probability that $\mathbf{s}$ takes on the value $\mathbf{x}_i \in X$ is given by
	\begin{equation}
	\Pr(\mathbf{s} = {\mathbf{x}_i}) = \prod_{m=1}^{a} \Pr(s_m = {\ell}_{i_m}).
	\label{eqn:9}
	\end{equation}

\subsection{Check Node Inputs}
 We use a $\kappa\times1$ column vector, {denoted by} $\mathbf{U}_{\left(\kappa\times 1\right)}^\mu$, to represent  the $\kappa$ single edge  LLRs input to the check nodes in $C(A)$ from the auxiliary variable nodes $A'$ at iteration $\mu$. 
	If $p$ decoding iterations are performed, all the single edge LLRs input to $C(A)$ at all iterations can be represented by the $\kappa \times p$ matrix $\mathbf{U}_{\left(\kappa\times p\right)} \triangleq [\mathbf{U}^1 \mathbf{U}^2 \cdots \mathbf{U}^p]$, where each element of  $\mathbf{U}$ can take one of  $t$  values. Therefore, $\mathbf{U}$ has $t^{{\kappa}p}$ possible realizations. We denote a given realization as $\mathbf{W}_k$ and the set of all possible realizations by $W = \left\{ \mathbf{W}_k| k = 1,2,\ldots,t^{{\kappa}p} \right\}$, where $|W|=t^{{\kappa}p}$ can be extremely large for practical values of $t$, $\kappa$, and $p$. As an illustration, $|W|\approx 2.3\times10^{105} $ for $t = 32$ (\textit{i.e.}, a 5-bit quantizer), $\kappa = 7$ check nodes {with external edges} connected to $A$, and $p = 10$ decoder iterations.
	
\section{Bounding the Error Probability of an Absorbing Set Decoder}\label{sec:Bound1}
For an absorbing set decoder operating on $\mathcal{D}(A)$ with independently chosen variable and check node inputs $\mathbf{s}$ (from the channel) and $\mathbf{U}$ (from outside $\mathcal{G}(A)$),  we define $\xi{(A)}$ to be the event that there remains  at least one bit error in $A$ after $p$ decoding iterations. The probability of error for an absorbing set decoder performing on $\mathcal{D}(A)$  can then be written  by conditioning the event $\xi{(A)}$ on all possible $\mathbf{s} ={\mathbf{x}_{i}} $ and $\mathbf{U} =  {\mathbf{W}_{k}} $ as follows:
	\begin{equation}
	\begin{split}
	\Pr(\xi{(A)}) &= \sum_{i=1}^{t^{ a}}\sum_{k=1}^{t^{{\kappa}p}}{  \Pr(\xi{(A)}|\mathbf{s} = {\mathbf{x}_{i}},\mathbf{U} = {\mathbf{W}_{k}})  } \\ 
	& \quad \cdot \Pr\left ( \mathbf{s} = {\mathbf{x}_{i}} ,\mathbf{U} =  {\mathbf{W}_{k}} \right ),
	\label{eqn:d5}
	\end{split} 
	\end{equation}
\noindent where $\Pr(\xi{(A)}|\mathbf{s} = {\mathbf{x}_{i}},\mathbf{U} = {\mathbf{W}_{k}})  $ is either $0$ or $1$, based on whether or not  the variable node input vector $\mathbf{s} = \mathbf{x}_{i}$ is  decoded  correctly after $p$ iterations when $\mathbf{U} = {\mathbf{W}_{k}} $ is the check node input matrix.
	
To help visualize~\eqref{eqn:d5}, we define a \emph{decodability array} for an absorbing set decoder, with $t^a$ columns corresponding to all possible variable node input  vectors $\mathbf{s} = \mathbf{x}_{i}$ and $t^{{\kappa}p}$ rows corresponding to all possible check node input matrices $\mathbf{U} = \mathbf{W}_{k}$. The columns are indexed by $a$-tuples over the set of quantizer levels, while the rows are indexed by $\kappa \times p$ matrices over the set of quantizer levels. We can then fill out the decodability array with
\begin{equation}
\mathbbm{1}_{(\mathbf{x}_i,\mathbf{W}_k )} 
  \triangleq \left\{\begin{matrix}
1,& \text{if the pair} \ (\mathbf{s} = {\mathbf{x}_{i}},\mathbf{U} = {\mathbf{W}_{k}}) \\ &\text{ is decoded incorrectly}, \\
0, & \text{if the pair} \ (\mathbf{s} = {\mathbf{x}_{i}},\mathbf{U} = {\mathbf{W}_{k}})\\& \text{ is decoded correctly}.
\end{matrix}\right.
\end{equation}
	  The resulting array is deterministic, \textit{i.e.}, it is not a function of the channel $\text{SNR}$. A pictorial representation of the decodability array is shown below:
	\begin{equation}
	\begin{blockarray}{cccccc}
	&\mathbf{x}_{1} & \cdots & \mathbf{x}_i & \cdots & \mathbf{x}_{t^a} \\
	\begin{block}{c(ccccc)}
	\mathbf{W}_{1} &  \mathbbm{1}_{(\mathbf{x}_{1},\mathbf{W}_{1} )} & \cdots & \mathbbm{1}_{(\mathbf{x}_i,\mathbf{W}_{1} )} & \cdots & \mathbbm{1}_{(\mathbf{x}_{t^a},\mathbf{W}_{1} )} \\
	\vdots & \vdots & \ddots & \vdots & \ddots & \vdots \\
	\mathbf{W}_k & \mathbbm{1}_{(\mathbf{x}_{1},\mathbf{W}_k )} & \cdots & \mathbbm{1}_{(\mathbf{x}_i,\mathbf{W}_k )} & \cdots & \mathbbm{1}_{(\mathbf{x}_{t^a},\mathbf{W}_k )} \\
	\vdots & \vdots & \ddots & \vdots & \ddots & \vdots \\
	\mathbf{W}_{t^{{\kappa}p}} & \mathbbm{1}_{(\mathbf{x}_{1},\mathbf{W}_{t^{{\kappa}p}} )} & \cdots & \mathbbm{1}_{(\mathbf{x}_i,\mathbf{W}_{t^{{\kappa}p}} )} & \cdots & \mathbbm{1}_{(\mathbf{x}_{t^a},\mathbf{W}_{t^{{\kappa}p}} )} \\
	\end{block}
	\end{blockarray}~.
	\label{eqn:dec1}
	\end{equation} 
	
	We now define the \emph{absorbing region}  of an absorbing set decoder as the set of all  pairs $(\mathbf{x}_{i},\mathbf{W}_{k})$ with `1' entries in the decodability array.\footnote{A related definition of an absorbing region was defined in~\cite{dolecek09}. {We note that, generally, the  decodability array can be constructed in this way for any problematic sub-graph and the corresponding ``absorbing region'' would refer to the portion of the array with `1' entries.}} Letting  $\psi$ represent the absorbing region, \textit{i.e.}, $ \psi = \left\{ \left(\mathbf{x}_i,\mathbf{W}_k\right)| \Pr(\xi{(A)}|\mathbf{s} = {\mathbf{x}_{i}},\mathbf{U} = {\mathbf{W}_{k}})  = 1\right\}$,   $\Pr(\xi{(A)})$ in~\eqref{eqn:d5} can be written as
	\begin{equation}
	\Pr(\xi{(A)}) = \sum_{\left(\mathbf{x}_i, \mathbf{W}_k\right) \in \psi}^{} \Pr\left ( \mathbf{s} = {\mathbf{x}_{i}} , \mathbf{U} =  {\mathbf{W}_{k}} \right ),
	\label{eqn:d51}
	\end{equation} 
	where~\eqref{eqn:7} and \eqref{eqn:9} indicate the dependence of $	\Pr(\xi{(A)})$ on SNR.
	Evaluating~\eqref{eqn:d51} is computationally complex, since the size of the decodability array ${t^{{\kappa}p}} \times t^{a}$ is  typically extremely large.
	In the rest of this section, we propose an approach to simplify the  problem of finding the probability $\Pr(\xi{(A)})$ of the absorbing region. 
	
	We proceed by proposing to lower bound $\Pr(\xi{(A)})$. {Assuming that $\mathbf{s}$ and $\mathbf{U}$ are chosen independently, }\eqref{eqn:d51}{ becomes}
	\begin{equation}
	\Pr(\xi{(A)})= \sum_{\left(\mathbf{x}_i, \mathbf{W}_k\right) \in \psi}^{}  \Pr\left ( \mathbf{s} = {\mathbf{x}_{i}} \right )\cdot \Pr\left (\mathbf{U} =  {\mathbf{W}_{k}} \right ),
		\label{eqn:newd51}
	\end{equation}
	{where we note that, in an absorbing set decoder, we are independently choosing an $\mathbf{s}$ and a $\mathbf{U}$, running the decoder to see if it is decoded incorrectly, which results in a ``1" in the decodability array, and then repeating this process for every possible combination in the array. After the process is complete, each entry in the array is either a ``1" or a ``0".} 
	
	{We now define the following sets,} which can be understood  by referring to  the  decodability array. First, for a given $\mathbf{W}_{k}$ (row of the decodability array), denote the set of all $\mathbf{x}_i$ (columns of the decodability array) for which the $ \left(\mathbf{x}_i,\mathbf{W}_k\right)$ pairs  cannot be decoded correctly as $\Psi\left(\mathbf{W}_{k}\right)$, \textit{i.e.}, $\Psi\left(\mathbf{W}_{k}\right) = \left\{ \mathbf{x}_i| \left(\mathbf{x}_i,\mathbf{W}_k\right) \in \psi \right\}$.
	This is equivalent to the set of all columns with entries `1' in a given row $\mathbf{W}_k$ of the decodability array.	
	Additionally, we let $\Psi(W)$ denote the set of all columns in the decodability array with  `1' entries  in every row, \textit{i.e.}, $ \Psi(W) = \left\{ \mathbf{x}_i|   \left(\mathbf{x}_i,\mathbf{W}_k\right) \in \psi, \forall~\mathbf{W}_{k}  \in W \right\}$, where we note that 
	\begin{equation}
	\Psi(W) = \bigcap_{k=1}^{t^{{\kappa}p}}\Psi\left(\mathbf{W}_{k}\right) .\label{eqn:dd6}
	\end{equation} 	
	In \eqref{eqn:newd51}, the error probability is a function of $\Pr(\mathbf{U} = {\mathbf{W}_{k}})$, which involves computing the probability of a particular set of $\kappa$ check node inputs (from outside $\mathcal{G}(A)$) for each of the $p$ iterations. If we are interested only in a  lower bound on the probability of $ \left(\mathbf{x}_i,\mathbf{W}_k\right)$ belonging to the absorbing region, this term can be eliminated from the calculation by  including in the sum only entries whose columns have a `1' in every row, \textit{i.e.}, the set $\Psi(W)$, which results in the following lower bound
	\begin{equation}
	\begin{split}
	\Pr(\xi{(A)}) &= \sum_{\left(\mathbf{x}_i, \mathbf{W}_k\right) \in \psi}^{} \Pr\left ( \mathbf{s} = {\mathbf{x}_{i}} \right )\cdot \Pr\left (\mathbf{U} =  {\mathbf{W}_{k}} \right ) \\ 
	&\geq  \sum_{\mathbf{x}_i \in \Psi(W), \mathbf{W}_k \in W }^{} \Pr\left ( \mathbf{s} = {\mathbf{x}_{i}} \right )\cdot \Pr\left (\mathbf{U} =  {\mathbf{W}_{k}} \right ) \\ 
	&= \sum_{\mathbf{x}_i \in \Psi(W)}^{} \Pr\left ( \mathbf{s} = {\mathbf{x}_{i}} \right )\sum_{\mathbf{W}_k \in W}^{} \Pr\left (\mathbf{U} =  {\mathbf{W}_{k}} \right ) \\ 
	&= \sum_{\mathbf{x}_i \in \Psi(W)}^{} \Pr\left ( \mathbf{s} = {\mathbf{x}_{i}} \right ).
	\label{eqn:d6}
	\end{split}
	\end{equation} 
	The lower bound in~\eqref{eqn:d6} implies that
	\begin{equation}
	\Pr(\xi{(A)})  \geq \lambda(A) \triangleq \sum_{\mathbf{x}_i \in \Psi(W)}^{} \Pr\left ( \mathbf{s} = {\mathbf{x}_{i}} \right ),
	\label{eqn:d7}
	\end{equation} 
	so that instead of including all the pairs in the decodability array with `1' entries, we only need to include  the columns with all `1' entries, which leads to the removal of the term $\Pr\left (\mathbf{U} =  {\mathbf{W}_{k}} \right ) $ from the expression for $	\Pr(\xi{(A)})$. This makes the evaluation of the lower bound in~\eqref{eqn:d7} dependent only on the absorbing set $A$ and not on the structure of the code containing $A$.\footnote{{If every column of the array has at least one ``0" entry, that means that every possible input to the ``absorbing set" can be decoded with some combination of check node inputs   
and we would obtain the trivial bound $\lambda = 0$; however, since such an object isn't problematic by our definition, a lower bound of zero makes sense.}}

	\section{Bounding the FER of an LDPC Code}\label{sec:bound2}
	In this section, we begin by deriving a lower bound on the FER of any LDPC code whose Tanner graph representation contains at least one instance of a given $(a,b)$ absorbing set $\mathcal{G}(A)$ in Section~\ref{da}. We then provide a series of approximations in Section~\ref{db} to reduce the complexity of evaluating the {bound.} 
{Finally,} in Section~\ref{dd} we provide some remarks concerning the {application, evaluation, and merits} of a code-independent bound on the FER of an LDPC code.
	
	\subsection{A Lower Bound on the FER of an LDPC Code}\label{da}
	We define ${{\mathcal{E}}}{(V)}$ as the event that there is at least one bit error in the set of variable nodes $V$ after  the quantized received vector $\mathbf{r}$ is decoded using a quantized decoder operating on the full code graph for $p$ iterations. Then the FER of the LDPC code can be written as  
	\begin{equation}
	\text{FER} = \Pr\left({{\mathcal{E}}}{(V)}\right) = \sum_{k=1}^{t^{ n}}{  \Pr({{\mathcal{E}}}{(V)}|\mathbf{r} = {\mathbf{z}_{k}})\cdot \Pr(\mathbf{r} = {\mathbf{z}_{k}})},
	\label{eqn:10}
	\end{equation}
	since there are $t^n$ possible realizations of $\mathbf{r}$. 
	
	Now let ${{\mathcal{E}}}{(A)}$ represent the event that there is at least one bit error in $\mathcal{G}(A)$ after $\mathbf{r}$ is decoded using a quantized  decoder operating on the full graph for $p$ iterations. Then $\Pr\left({{\mathcal{E}}}{(A)}\right)$ determines the contribution of  $A$ to the FER, and we can therefore write 
	\begin{equation}
	\text{FER} \geq \Pr\left({{\mathcal{E}}}{(A)}\right) = \sum_{k=1}^{t^{ n}}{  \Pr({{\mathcal{E}}}{(A)}|\mathbf{r} = {\mathbf{z}_{k}})\cdot \Pr(\mathbf{r} = {\mathbf{z}_{k}})}.\label{eqn:d22}
	\end{equation}
	 Now we make the important observation that, since $\Pr({{\mathcal{E}}}(A))$ depends only on the input vector $\mathbf{s}$ received by the  variable nodes of $A$ and the input matrix $\mathbf{U}$ received by the check nodes connected to $A$ during the $p$ iterations of decoding,
	it can also  be written as
	\begin{equation}
	\begin{split}
	\Pr({{\mathcal{E}}}{(A)}) = \sum_{i=1}^{t^{ a}}\sum_{k=1}^{t^{n-a}}{  \Pr({{\mathcal{E}}}{(A)}|\mathbf{s} = {\mathbf{x}_{i}},\mathbf{U} = {\mathbf{W}_{k}})  } \\ 
	\cdot \Pr\left ( \mathbf{s} = {\mathbf{x}_{i}},\mathbf{U} =  {\mathbf{W}_{k}} \right ),
	\end{split}
	\label{eqn:x1}
	\end{equation}
	 where we note that $\Pr({{\mathcal{E}}}{(A)})$ represents the probability that $A$ is in error for a \emph{full graph decoder}, whereas $\Pr(\xi{(A)})$ in~\eqref{eqn:d5} represents the probability that $A$ is in error for an absorbing set decoder. Here, {unlike in~}\eqref{eqn:newd51}, $\mathbf{s}$ and $\mathbf{U}$ are dependent variables, since the absorbing set check node input matrix $\mathbf{U}$ depends on the variable node input vector $\mathbf{s}$ in a full graph decoder. We can now state the  following theorem.
	\begin{theo}
		For any LDPC code containing the absorbing set $\mathcal{G}(A)$, $\lambda(A)$ defined in~\eqref{eqn:d7} lower bounds $\Pr\left({{\mathcal{E}}}{(A)}\right)$, \textit{i.e.},
		\begin{equation}
			\Pr\left({{\mathcal{E}}}{(A)}\right) \geq \lambda(A).
			\end{equation}  
	\end{theo}
	\textit{Proof}: We begin by defining a decodability array for $\mathcal{G}(A)$, similar to~\eqref{eqn:dec1}, but for a full graph decoder. In this case, however, each of the $t^a$ columns, representing a given variable node input vector {($a$-tuple)} $\mathbf{s} = \mathbf{x}_i$  in $A$, contains entries in at most $t^{n-a}$ rows, one for each of the $t^{n-a}$ possible check node input matrices $\mathbf{U} = \mathbf{W}_k$ that result from using the full graph decoder to decode $\mathbf{s} = \mathbf{x}_i$ combined with one of the $t^{n-a}$ input vectors in $V\setminus A$, where we note that some of the $t^{n-a}$ decoder results may give the same check node input matrix. Also, some of the entries in the decodability array may be \emph{blank}, corresponding to cases where the full graph decoder never results in a particular combination of $\mathbf{s} = \mathbf{x}_i$ and $\mathbf{U} = \mathbf{W}_k$. 
	
	We next fill in the non-blank entries in the decodability array according to whether the pair $(\mathbf{s} = \mathbf{x}_i, \mathbf{U} = \mathbf{W}_k)$ is decoded correctly (a `0') or incorrectly (a `1') by the full graph decoder. We now define the absorbing set region of a full graph decoder as the set of all pairs $(\mathbf{x}_i,\mathbf{W}_k)$ with `1' entries in the decodability array and denote it as $ \psi' = \left\{ \left(\mathbf{x}_i,\mathbf{W}_k\right)| \Pr({{\mathcal{E}}}{(A)}|\mathbf{s} = {\mathbf{x}_{i}},\mathbf{U} = {\mathbf{W}_{k}})  = 1\right\}$. We can then express~\eqref{eqn:x1} in terms of this absorbing set region as
		\begin{equation}
	\Pr({{\mathcal{E}}}{(A)}) = \sum_{\left(\mathbf{x}_i, \mathbf{W}_k\right) \in \psi'}^{} \Pr\left ( \mathbf{s} = {\mathbf{x}_{i}} , \mathbf{U} =  {\mathbf{W}_{k}} \right ).
	\label{eqn:d23}
	\end{equation} 
	Further, let $\Psi'(W)$ be the set of all columns in the decodability array with \emph{either} `1' or blank entries in every row, \textit{i.e.}, the set of all variable node input vectors {$\mathbf{s} = \mathbf{x}_i$ that} are not decoded correctly by the full graph decoder. We can now write
		\begin{equation}
	\begin{split}
	\Pr({{\mathcal{E}}}{(A)}) &= \sum_{\left(\mathbf{x}_i, \mathbf{W}_k\right) \in \psi'}^{} \Pr\left ( \mathbf{s} = {\mathbf{x}_{i}} ,\mathbf{U} =  {\mathbf{W}_{k}} \right ) \\ 
	&\geq  \sum_{\mathbf{x}_i \in \Psi'(W), \mathbf{W}_k \in W }^{} \Pr\left ( \mathbf{s} = {\mathbf{x}_{i}} ,\mathbf{U} =  {\mathbf{W}_{k}} \right ) \\ 
	&= \sum_{\mathbf{x}_i \in \Psi'(W)}^{} \Pr\left ( \mathbf{s} = {\mathbf{x}_{i}} \right ).
	\label{eqn:x3}
	\end{split}
	\end{equation} 
	An  important observation now follows: if a column $\mathbf{s} = \mathbf{x}_i$ in the decodability array for the \textit{absorbing set decoder} contains all `1' entries, \textit{i.e.}, if $\mathbf{x}_i \in \Psi(W)$, then it must contain either `1' or blank entries in every row of the decodability array for the \textit{full graph decoder}, \textit{i.e.}, $\mathbf{x}_i \in \Psi'(W)$. {Note, however, that the converse is not true.} In other words, if $\mathbf{x}_i \in \Psi'(W)$, it does not follow that $\mathbf{x}_i \in \Psi(W)$, since blank entries in the decodability array for the full graph decoder (corresponding to check node input matrices that never occur) could be decoded correctly by the absorbing set decoder.
	
	Now defining
		\begin{equation}
	\lambda'(A) \triangleq \sum_{\mathbf{x}_i \in \Psi'(W)}^{} \Pr\left ( \mathbf{s} = {\mathbf{x}_{i}} \right ),
	\label{eqn:d25}
	\end{equation} 
	it follows that
		\begin{equation}
		\begin{split}
\lambda'(A) \triangleq &\sum_{\mathbf{x}_i \in \Psi'(W)}^{} \Pr\left ( \mathbf{s} = {\mathbf{x}_{i}} \right ) \\&\geq  \sum_{\mathbf{x}_i \in \Psi(W)}^{} \Pr\left ( \mathbf{s} = {\mathbf{x}_{i}} \right ) = \lambda(A),
	\label{eqn:d26}
	\end{split}
	\end{equation} 
	and
		\begin{equation}
	\Pr({{\mathcal{E}}}(A)) \geq \lambda'(A) \geq \lambda(A).
	\label{eqn:d27}
	\end{equation} 
	\hfill $ \Box$
	
If there are $N$ occurrences of an $(a,b)$ absorbing set, denoted by $\mathcal{G}(A_i)$, $i = 1,2,\ldots,N$, in a given code, the contribution of all $N$ absorbing sets of this type to the FER is given by
	$\Pr{  \left({\bigcup_{i=1}^{N}{{\mathcal{E}}}{(A_i)}}\right)  }$.
	Since these absorbing sets may not be the only cause of decoding errors, $\Pr{  \left({\bigcup_{i=1}^{N}{{\mathcal{E}}}{(A_i)}}\right)  }$ gives a lower bound  on the FER, \textit{i.e.}, 
	\begin{equation}
	\text{FER} \geq \Pr{  \left({\bigcup_{i=1}^{N}{{\mathcal{E}}}{(A_i)}}\right)  } .
	\label{eqn:12}
	\end{equation}
		Assuming that all $(a,b)$ absorbing sets within a given code have the same $\Pr{  \left({{{\mathcal{E}}}{(A_i)}}\right) }$, {denoted by} $\Pr{  \left({{{\mathcal{E}}}{(A)}}\right) }$,\footnote{This assumption is based on the symmetry of the channel and is particularly relevant for the structured codes, due to their additional symmetry. A similar assumption is made in~\cite{dolecek09,schelegel13,butler14}.} an immediate result of~\eqref{eqn:12} is that
	\begin{equation}
	\text{FER} \geq  \Pr{  ({{\mathcal{E}}}{(A)})  } \geq \lambda(A).
	\label{eqn:12a}
	\end{equation}
	Furthermore, since 
	\begin{equation}
	\begin{split}
	\Pr{  \left({\bigcup_{i=1}^{N}{{\mathcal{E}}}{(A_i)}}\right)  } \geq &\sum_{i=1}^{N}\Pr{  \left({{{\mathcal{E}}}{(A_i)}}\right)  }\\ &- \sum_{1 \leq i < j \leq N}^{}\Pr{  \left({{{\mathcal{E}}}{(A_i)} \cap {{\mathcal{E}}}{(A_j)}}\right)  },
	\label{eqn:13}
	\end{split}
	\end{equation}
	\eqref{eqn:12} and \eqref{eqn:13} can be combined to give the following lower bound
	\begin{equation}
	\text{FER} \geq \sum_{i=1}^{N}\Pr{  \left({{{\mathcal{E}}}{(A_i)}}\right)  } - \sum_{1 \leq i < j \leq N}^{}\Pr{  \left({{{\mathcal{E}}}{(A_i)} \cap {{\mathcal{E}}}{(A_j)}}\right)  }.
	\label{eqn:14}
	\end{equation}
 We now assume that any two error events ${{\mathcal{E}}}{(A_i)}$ and ${{\mathcal{E}}}{(A_j)}$ associated with the same $(a,b)$ absorbing set are independent, \textit{i.e.},
	\begin{equation}
	\Pr{  \left({{{\mathcal{E}}}{(A_i)} \cap {{\mathcal{E}}}{(A_j)}}\right)} =  \Pr{  \left({{\mathcal{E}}}{(A_i)} \right)} \cdot \Pr{  \left({{\mathcal{E}}}{(A_j)} \right)}.
	\label{eqn:15}
	\end{equation} 
	This assumption is made for simplicity and is based on the observation  that most pairs of a given absorbing set appearing in a code are disjoint, in the sense that they do not have any nodes in common.
	Using this assumption, the right hand side of \eqref{eqn:14} can be written as
	\begin{equation}
	N\Pr{  \left({{{\mathcal{E}}}{(A)}}\right)  } - \binom{N}{2}\left( \Pr{  \left({{\mathcal{E}}}{(A)} \right)}\right)^2.
	\label{eqn:16}
	\end{equation} 
	Further, as noted in~\cite{dolecek09}, the fact that the channel LLRs in the error floor region  are typically  large implies  that the chance of more than one absorbing set $\mathcal{G}(A)$ receiving low channel LLRs, and thus causing decoding errors, is small. This, combined with the fact that  the second term in  \eqref{eqn:16} will not have a significant impact (since $\Pr  ({{\mathcal{E}}}{(A)})$ will be small and thus $(\Pr  ({{{\mathcal{E}}}{(A)}}))^2 \ll \Pr  ({{{\mathcal{E}}}{(A)}})$ in the error floor) and can thus be neglected, results in the following approximate lower bound on the FER in the error floor region of an LDPC code containing $N$ instances of the absorbing set $\mathcal{G}(A)$: 
	\begin{equation}
	\text{FER} \gtrapprox N\Pr{  \left({{{\mathcal{E}}}{(A)}}\right)  } \geq N\lambda(A),
	\label{eqn:16a}
	\end{equation} 
 where the accuracy of the approximate bound in~\eqref{eqn:16a} depends on the tightness of the bound in~\eqref{eqn:d7}. Furthermore, if $\mathcal{G}(A)$ is the \emph{most harmful} or \emph{dominant} absorbing set in a code, $N\lambda(A)$ represents an estimate of its FER performance in the error floor region.\footnote{{In the case where more than one absorbing set is believed to be dominant, the maximum of all the lower bounds can be used to form an error estimate.}}
	
	Expressions~\eqref{eqn:12a} and~\eqref{eqn:16a} represent  a true lower bound and an approximate lower bound, respectively, valid in the error floor region, in terms of $\lambda(A)$, defined in  \eqref{eqn:d7}. 
	The multiplicities of the different absorbing sets needed to evaluate~\eqref{eqn:16a} may be derived either using analytical or semi-analytical methods, such as those given in~\cite{dolecek10,Liu10,yoones16},

	\subsection{Approximating the Lower Bound on FER}\label{db}
	In this section, we propose a reduced complexity method to approximate $\lambda(A)$. Although  the term $\Pr\left (\mathbf{U} =  {\mathbf{W}_{k}} \right ) $ was eliminated from the expression for $\Pr(\xi{(A)})$ in \eqref{eqn:d6}, thus making the lower bound code-independent and simplifying the expression, calculating $\lambda(A)$ in~\eqref{eqn:d7} still depends on finding $\Psi(W)$, which, in-turn requires examining all $\mathbf{W}_k \in W$ as shown in~\eqref{eqn:dd6}. In other words, all $t^{{\kappa}p}$ rows of the decodability array should be examined for each of the $t^a$ columns  $\mathbf{x}_i$. 
	Therefore, instead of finding $\Psi(W)$, we consider the less computationally complex set 
	\begin{equation} {\hat\Psi(W)} =  \bigcap_{m=1}^{{M}}\Psi\left(\mathbf{W}_{k_m}\right),k_1,k_2,\ldots,k_m\in\{1,2,\ldots,t^{{\kappa}p}\},
	\label{eqn:d81}
	\end{equation} 
	which involves examining only a subset of $M$ rows of the decodability array. By properly choosing the $M$ rows and finding the columns with all `1' entries in these rows, it is possible to obtain a good  approximation to the set of columns with `1' entries in every row, allowing us to compute 
	\begin{equation}
	\hat{\lambda}(A) \triangleq  \sum_{\mathbf{x}_i \in {\hat\Psi(W)}}^{} \Pr\left ( \mathbf{s} = {\mathbf{x}_{i}}  \right ) \approx \lambda(A),
	\label{eqn:d8}
	\end{equation} 
	which results in the \emph{approximate lower bound}\footnote{We use this term to emphasize the fact that approximations are used in calculating $\lambda(A)$.} 
	\begin{equation}
	\text{FER} \gtrapprox N\hat{\lambda}(A).
	\label{eqn:d8a}
	\end{equation}

	In the following, we explain how the approximate lower bound $\hat{\lambda}(A)$ is  calculated.
	We first assume that $M$ rows of the decodability array, {denoted by} $\mathbf{W}_{k_m}$ for $m = 1, 2, \ldots, M$, have been selected. The calculation of~\eqref{eqn:d8} then involves two steps:
	\begin{enumerate}
		\item Finding the set $\hat{\Psi}(W)$. This is achieved by operating the absorbing set decoder on $A$ for each $(\mathbf{s} = {\mathbf{x}_{i}},\mathbf{U} = {\mathbf{W}_{k_m}})$. Then, using~\eqref{eqn:d81}, if the decoder fails to correctly decode $\mathbf{s} = \mathbf{x}_{i}$ for all the $\mathbf{W}_{k_m}, m = 1,2,\ldots, M$ , it follows that $\mathbf{s} = \mathbf{x}_{i}\in\hat{\Psi}(W)$. Otherwise, $\mathbf{s} = \mathbf{x}_{i}$ is discarded.
				\item Summing the $\Pr\left ( \mathbf{s} = {\mathbf{x}_{i}} \right )$ for all $\mathbf{s} = \mathbf{x}_{i}\in\hat{\Psi}(W)$, where $\Pr\left ( \mathbf{s} = {\mathbf{x}_{i}} \right )$ is obtained using  ~\eqref{eqn:7} and~\eqref{eqn:9}.
	\end{enumerate}
	
	%
	
	In order to obtain a computationally efficient approximation, we should choose  rows expected to have a small number of  `1's, since they eliminate more columns than rows with a large number of `1's . In other words, the $M$ rows should be chosen as a set of check node input matrices $\mathbf{U} = \mathbf{W}_{k_m}$ that we expect to result in a small number of input vectors to the absorbing set $\mathbf{s} = \mathbf{x}_i$ that cannot be decoded correctly. Rows which we expect will lead to incorrect decoding of most input vectors $\mathbf{s} =  \mathbf{x}_i$, on the other hand, are not useful. Therefore, we try to avoid such rows. Before proceeding, we review some important facts regarding the dynamics of absorbing sets in the high SNR region (with highly reliable input channel values). For such absorbing sets, after a certain number of iterations, it is common for the LLRs received by the check nodes in $C(A)$ from the variable nodes in $V \setminus A$ to grow rapidly and reach the maximum quantizer level $\ell_t$ (or the saturation level) within a few iterations~\cite{dol09}. For example, the analysis in \cite{tomasoni17} starts from the point where  all the LLRs have already converged to $\ell_t$. This motivates our choice of \emph{Row Set I}, where we consider only the row $\mathbf{U} = \mathbf{W}_{\max}$, where $\mathbf{W}_{\max} \triangleq [\ell_t]_{\kappa\times p}$. Further, in~\cite{schelegel13,butler14}, and~\cite{angarita14} it is stated that slowing down convergence to the maximum level for the LLRs inside an absorbing set often leads to an increase in the probability of correct decoding. This motivates our choice of \emph{Row Set II}, where we consider a more gradual increase of the input LLRs to the check nodes in $C(A)$, which corresponds to choosing another row in the decodability array. We also  propose \emph{Row Set III}, which combines Row Sets I and II  using~\eqref{eqn:d81}. In general, we have found that rows $\mathbf{W}_{k_m}$ with a high probability of correct decoding have increasing LLRs with iterations, and no negative LLRs (assuming all-zero transmission), a point also noted in~\cite{schelegel13}.

\subsubsection{Row Set I}
The first candidate check node input matrix that we consider is $\mathbf{U} = \mathbf{W}_{\max}$, which is based on the following assumption.
\begin{assum} \label{as:2}
		\textit{For any  absorbing set input vector $\mathbf{s} = \mathbf{x}_i$ such that $\left( \mathbf{x}_{i},\mathbf{W}_{\max}\right) \in \psi$, \textit{i.e.}, $\left(\mathbf{x}_i,\mathbf{W}_{\max}\right)$ cannot be decoded correctly, any other input pair $\left(\mathbf{x}_i,\mathbf{W}_{k'}\right)$  also cannot be decoded correctly, \textit{i.e.},
			\begin{equation}
			\Psi\left(\mathbf{W}_{\max}\right) \subseteq \Psi\left(\mathbf{W}_{k'}\right), \forall ~\mathbf{W}_{k'} \neq \mathbf{W}_{\max} .
			\label{eqn:d125}
			\end{equation} }
\end{assum}

In other words, it is assumed that, if the row in the decodability array associated with $\mathbf{W}_{\max}$ has a `1'  entry in a column, the remaining rows will also have `1' entries in the same column.
This assumption is based on the behavior of maximum likelihood (ML) decoding, where the best performance is achieved if the distance from the received vector $\mathbf{r}$ to the most reliable channel LLR vector is minimum, \textit{i.e.}, all the variable nodes in $V\setminus A$ receive the maximum quantizer output $\ell_{{t}}$. Extending this logic to an iterative MP decoder would imply that, if the variable nodes in $V\setminus A$ send $\mathbf{U} = \mathbf{W}_{\max}$ to the check nodes in $C(A)$, the best decoding performance is achieved. Although this is not necessarily the case for iterative decoders, this assumption along with similar earlier arguments from~\cite{tomasoni17},~\cite{bani13}, and~\cite{dol09} motivates choosing
	\begin{equation}
	\hat\Psi{(W)} =   \Psi\left(\mathbf{W}_{\max}\right) .
	\label{eqn:412}
	\end{equation}
Note that this choice of $\hat\Psi{(W)}$ yields a $\hat{\lambda}(A)$ that is significantly less complex to calculate  than $\lambda(A)$, since $\hat{\lambda}(A)$ requires examining only one row $\mathbf{W}_{\max}$, while  $t^{{\kappa}p}$ rows must be examined to calculate $\lambda(A)$.
	
\begin{figure}[t]
		\centering
		\includegraphics[width=\columnwidth]{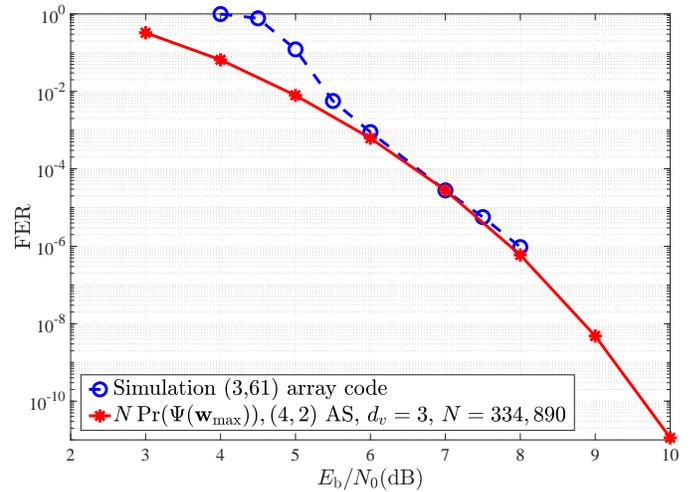}\\
		\caption{ Approximate lower bound of~\eqref{eqn:d8a} based on a $(4,2)$ absorbing set for a $\left(3,61\right)$ array code with a $\text{Q}_{3.2}$ uniform quantizer with $\Phi(0) = 4.25$ and an SPA decoder using Row Set I.}\label{1a}
	\end{figure}

Fig.~\ref{1a} shows the approximate lower bound of~\eqref{eqn:d8a} based on \eqref{eqn:412} for a $(4,2)$ absorbing set in a $(3,61)$ array code~\cite{fan} of length $n = 3721$ and rate $R = 0.9514$ with a $\text{Q}_{3.2}$ uniform quantizer and an SPA decoder, where a multiplicity of 334,890 was assigned to the absorbing set (see~\cite{dolecek10}). This absorbing set was chosen because it was shown to be the dominant one for $(3,K)$ array codes with an SPA decoder and a $\text{Q}_{3.2}$ uniform quantizer~\cite{dol09}.\footnote{We remark again that this (4,2) absorbing set can also be considered as an ETS or LETS.} 
	The FER of the simulated $(3,61)$ array code is also shown for comparison. 
	We observe that the approximate lower bound closely follows the simulated performance in the error floor region, thus supporting the choice of $\mathbf{U} = \mathbf{W}_{\max}$.
	
	\subsubsection{Row Set II}
\begin{figure}[t]
		\centering
		\includegraphics[width=0.6\columnwidth]{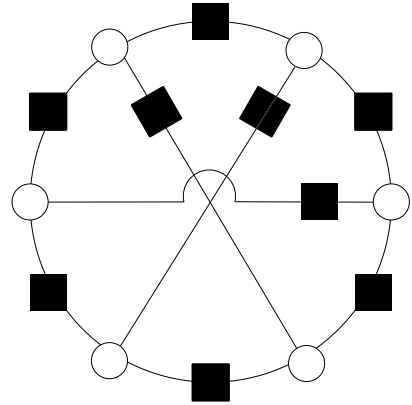}
		\caption{An illustration of a $\left(6,0\right)$ absorbing set with girth 8.}\label{6_0}
	\end{figure}

	The results of~\cite{siegel14} indicate that for certain LDPC codes, the MSA decoder with a quasi-uniform quantizer can have error floor performance very close to that of an unquantized SPA decoder. As an example, for $(3,K)$ array codes decoded using the MSA  decoder with a 5-bit quasi-uniform quantizer, we find that the dominant error patterns are $\left(6,0\right)$ absorbing sets with girth $8$, as shown in Fig.~\ref{6_0}. This absorbing set is the support of a codeword and represents the minimum distance of these codes (which illustrates the efficiency of the quasi-uniform quantizer for MSA decoding).\footnote{This (6,0) absorbing set, or codeword, can also be classified as an ETS or LETS.} When we apply Assumption~\ref{as:2} to the $\left(6,0\right)$ absorbing set with different quasi-uniform quantizers and use the multiplicity of the $\left(6,0\right)$ absorbing set from~\cite{Liu10},  however, we find that the approximate lower bound is, in fact, larger than the associated simulation result for the MSA decoder. In other words, 	our results in this case show that there must exist columns in the decodability array with `1' entries in the row associated with $\mathbf{W}_{\max}$ but `0' entries in another row. This is consistent with the results of~\cite{schelegel13,butler14, angarita14,isit19}, \textit{i.e.}, that slowing down the convergence of the LLRs inside the absorbing set can increase the probability of correct decoding. Therefore, we conclude that Assumption 1, which is based on ML decoders, is not necessary valid for all MP decoders. This suggests choosing $\mathbf{U} = \mathbf{W}'$, where $\mathbf{W}'$ is a check node input matrix, corresponding to some other row of the decodability array, that can lead to correct decoding of some absorbing set input vectors $\mathbf{s} = \mathbf{x}_i$ when $\mathbf{U} = \mathbf{W}_{\max}$ does not lead to correct decoding.
	In~\cite{sun04,schelegel13,butler14}, the authors model the dynamics of an absorbing set by applying Density Evolution (DE) to the messages coming from outside the absorbing set, where a Gaussian distribution for the LLRs received by the check nodes in $C(A)$ at each iteration is assumed. These distributions are represented by their mean and variance, which are shown to be increasing with iteration number. Here, we make use of those results and extend them to our code-independent framework by considering a check node input matrix $\mathbf{U} = \mathbf{W}_{\text{inc}} $ for which the  LLRs increase gradually until reaching the maximum quantizer levels, thereby slowing down the convergence speed of the LLRs passed along the edges of the absorbing set decoder. To this end,  the elements of $\mathbf{U}_{\left(\kappa\times 1\right)}^1$ are set equally to the lowest positive quantizer level. We then let this value increase with the iteration number $\mu$, so that  each of the  $\frac{t}{2}$ positive quantizer levels is used $h$ times before moving to the next larger level, for a total of $p = h \left(\frac{t}{2}\right)$ iterations, resulting in the check node input matrix\footnote{We do not use negative quantizer  values because we are interested in rows that can decode most of the input patterns, and check node input matrices with negative values typically have a small probability of correct decoding  (\textit{e.g.}, see~\cite{siegel14}).}
	\begin{equation}
	\mathbf{U} = \mathbf{W}_{\text{inc}} = \left[\left[{\ell}_{\frac{t}{2}+1} \right]_{\left(\kappa\times h\right)} \ldots \left[{\ell}_{t} \right]_{\left(\kappa\times h\right)}      \right]_{\left(\kappa\times p\right)},
	\label{eqn:i3}
	\end{equation}
	and the set
	\begin{equation}
	\hat\Psi{(W)} =   \Psi\left(\mathbf{W}_{\text{inc}}\right).
	\label{eqn:i3b}
	\end{equation}
As in the case of Row Set I, $\hat{\lambda}(A)$ is significantly less complex to calculate  than $\lambda(A)$.
	The choice of $h$ and the general trajectory of the increasing quantizer levels give us some options for choosing $\mathbf{U} = \mathbf{W}_{\text{inc}}$. According to our experience, for a $(6,0)$ absorbing set with a 5-bit quasi-uniform quantizer, increasing $h$ beyond $3$ did not improve the approximate lower bound based on $\mathbf{U} = \mathbf{W}_{\text{inc}}$.

	\subsubsection{Row Set III}
	Finally, we can apply~\eqref{eqn:d81} to the two proposed sets $\Psi\left(\mathbf{W}_{\text{inc}}\right)$ and $\Psi\left(\mathbf{W}_{\max}\right)$ to obtain
	\begin{equation}
	\hat{\Psi}(W) =    \Psi(\mathbf{W}_{\text{inc}})  \cap \Psi\left(\mathbf{W}_{\max}\right).
	\label{eqn:i4}
	\end{equation}
	$\hat\Psi{(W)}$ again yields a $\hat{\lambda}(A)$ that is significantly less complex to calculate  than $\lambda(A)$. The procedure to find the proposed $\hat\Psi{(W)}$ is described in Algorithm~\ref{alg4}.
	\begin{algorithm}[t]
		\caption{Calculate $\hat{\lambda}(A) \triangleq  \sum_{\mathbf{x}_i \in {\hat\Psi(W)}}^{} \Pr\left ( \mathbf{s} = {\mathbf{x}_{i}} \right )$ }
		\begin{algorithmic}[1]\label{alg4}
			\STATE $\hat\Psi{(W)} \gets \varnothing $ (the empty set)
			\FORALL {$\mathbf{x}_i \in X$}
			\STATE The absorbing set decoder tries to decode ${\mathbf{x}_{i}}$ with $\mathbf{U} = \mathbf{W}_{\max}$;
			\IF {the absorbing set decoder fails}
			\STATE The absorbing set decoder tries to decode ${\mathbf{x}_{i}}$ with $\mathbf{U} = \mathbf{W}_{\text{inc}}$;
			\IF {the absorbing set decoder fails}
			\STATE  $\hat\Psi{(W)} = \hat\Psi{(W)} \cup \mathbf{x}_i$;
			\ENDIF
			\ENDIF
			\ENDFOR	
				\RETURN
		\end{algorithmic}
	\end{algorithm}

\begin{figure}[t]
	\centering
	\includegraphics[width=\columnwidth]{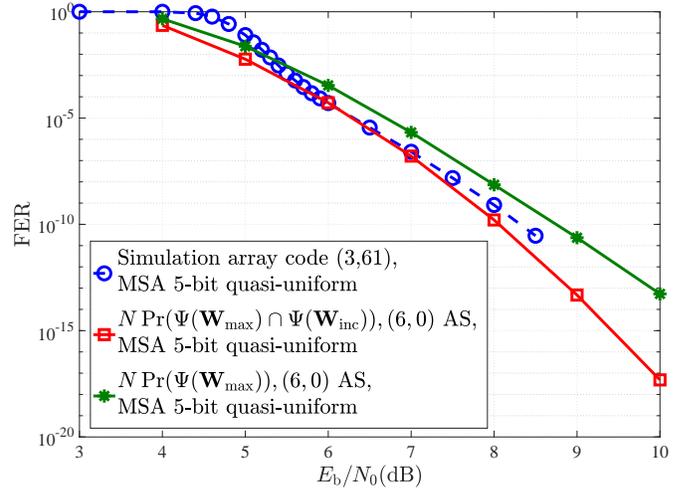}\\
	\caption{Approximate lower bound of ~\eqref{eqn:d8a} based on Row Sets I and III based on a $\left(6,0\right)$ absorbing set in a $\left(3,61\right)$ array code with a 5-bit quasi-uniform quantizer and an MSA decoder ($N = 2,195,390$).}\label{61_5}
\end{figure}

	As noted previously, the calculation of $\hat{\lambda}(A)$  can be seen as a two-step process: finding the set $\hat\Psi{(W)}$ by operating the absorbing set decoder and then calculating the probability of $\hat\Psi{(W)}$ using~\eqref{eqn:d8}. 
	In  Fig.~\ref{61_5}, for the $\left(6,0\right)$ absorbing set, the approximate lower bound of~\eqref{eqn:d8a} based on the set $\hat{\Psi}(W) =    \Psi(\mathbf{W}_{\text{inc}})  \cap \Psi\left(\mathbf{W}_{\max}\right)$, the bound  based only on the set $\Psi\left(\mathbf{W}_{\max}\right)$, and the simulated performance  are shown for a $\left(3,61\right)$ array code~\cite{fan} with a  $5$-bit quasi-uniform quantizer and an MSA decoder. We observe that in this case the approximate lower bound based on Row Set III gives a better result than the one obtained using only  $\Psi\left(\mathbf{W}_{\max}\right)$, \textit{i.e.}, Row Set I. It is worth noting that, to reduce the complexity of applying Algorithm~\ref{alg4}, we start with $\mathbf{W}_{\max}$, since it is likely to eliminate the most input vectors $\mathbf{s} = \mathbf{x}_i$. Then we look for other rows that might succeed where $\mathbf{W}_{\max}$ fails, so that, after checking $\mathbf{W}_{\max}$, it is only necessary to run the absorbing set decoder for those  $\mathbf{x}_i$'s with a `1' in the row  of the decodability array associated with $\mathbf{W}_{\max}$. 
	
	\subsection{Remarks}\label{dd}
	Due to its generality and simplicity, the code-independent  approximate lower bound on the FER in~\eqref{eqn:d8a} is a useful tool in predicting the high SNR performance of quantized LDPC decoders based on the presence of a given absorbing set (or general problematic sub-graph). Below we summarize this concept and pinpoint its strengths.
	\begin{itemize}
		
		\item {\emph{Application:}}  The lower bound $\lambda(A)$ indicates that any code containing at least one instance of a given absorbing set $A$ cannot achieve an FER lower than that value. This statement, although not strictly true for the approximate lower bound $\hat{\lambda}(A)$, can loosely be considered to have the same implication, as our numerical results show in the next section. In the same fashion, given that the multiplicity of the absorbing set is $N$, the approximate bound indicates that one cannot achieve an FER lower than $N\hat{\lambda}(A)$. In the case that $A$ is the dominant absorbing set in a code, the approximate lower bound $N\hat{\lambda}(A)$ becomes an estimate of its FER performance. Since ${\lambda}(A)$ and its approximation $\hat{\lambda}(A)$ are based only on an absorbing set, rather than a specific code, ~\eqref{eqn:12a},~\eqref{eqn:16a}, and~\eqref{eqn:d8a} apply to any code containing that absorbing set.

		\item {\emph{Complexity:}} An advantage of the code-independent bound is that its computational complexity relative to similar code-dependent methods, such as the error floor approximation of~\cite{dolecek10}, is only on the order of ${a}/{n}$, since the evaluation of $\hat{\Psi}(W)$ is  performed solely on the absorbing set of length $a$ and not on the entire code of length $n$. For example, for the $\left(3,61\right)$ array code and the $\left(6,0\right)$ absorbing set, the complexity of the code independent bound is only is $0.0016$ times that of the code-dependent bound.\footnote{This is true under identical conditions, such as dividing the variable nodes into two groups as proposed in~\cite{dolecek09}. } Furthermore, the time needed to evaluate the code-independent approximate lower bound is much less than for  Monte Carlo simulation for small values of the FER. For example, for the $(3,61)$ array code and the $(6,0)$ absorbing set, an  FER of only $10^{-4}$ can be achieved with Monte Carlo simulation in the time needed to evaluate the code-independent bound, which can accurately predict performance at FERs many orders of magnitude lower.

	\item {\emph{Code rate dependency:} Assuming the BPSK mapping described in Section} \ref{sec:mapping}{, calculating  $\hat{\lambda}(A)$ in }\eqref{eqn:d8} {requires the probability distribution given in} \eqref{eqn:7}{ and thus the bound is a function of the channel noise parameter $\sigma$. It follows that}
	\begin{equation}
	\hat{\lambda}(A) = f\left( R, \frac{E_\text{b}}{N_\text{0}} \right) = f\left(1, \frac{E'_\text{b}}{N_\text{0}}\right) = f\left(1, R\frac{E_\text{b}}{N_\text{0}}\right),
	\label{eqn:214a}
	\end{equation}
	{where $E_\text{b}' \triangleq R E_\text{b}$, or, expressing $ R{E_\text{b}}/{N_\text{0}}$ in decibels,}
	\begin{equation}
	\begin{split}
	\hat{\lambda}(A) &= f\left(1, R\frac{E_\text{b}}{N_\text{0}}(\text{dB})\right) \\&=  f\left(1, 
	\frac{E_\text{b}}{N_\text{0}}(\text{dB})+ 10\log(R)\right),
	\label{eqn:214}
	\end{split}
	\end{equation}
	{\textit{i.e.}, the approximate lower bound on FER in }\eqref{eqn:d8a}{ for code rate $R$ can be expressed as a  function of the bound for code rate $R' = 1$. The significance of}~\eqref{eqn:214} {is that it is only necessary to derive $\hat{\lambda}(A)$ for rate $R'=1$, which can then be used to derive $\hat{\lambda}(A)$ for  any rate, where the code rate $R$ only shifts the FER vs. ${E_\text{b}}/{N_\text{0}} (\text{dB})$ curve by $10\log(R) (\text{dB})$.}

		\item {\emph{SNR dependency:}} In the calculation of the  bound,  the set  $\hat{\Psi}(W)$ does not depend on the channel SNR. (Only the probabilities of the members of   $\hat{\Psi}(W)$ depend on SNR.) Therefore, the simulations needed to find the sets comprising $\hat{\Psi}(W)$ must be performed only once, and only a simple analysis is required to evaluate the bound once  $\hat{\Psi}(W)$  has been found, resulting in an easy way to estimate high SNR performance. On the other hand, in Monte Carlo simulations of a code, each SNR must be simulated separately, which is particularly time-consuming at high SNR. 
		
		\item {\emph{Generality}: A primary motivation of this work was to try to determine the impact that the existence of certain absorbing sets has on the performance of quantized LDPC decoders. This impact is characterized explicitly by the lower bounds }\eqref{eqn:12a},~\eqref{eqn:16a}, and~\eqref{eqn:d8a}{. For many classes of LDPC codes and decoders, the problematic objects are well-known and have been enumerated (see, e.g., }\cite{richardson03}-\cite{LETS}{). Moreover, a comprehensive ontology of possible trapping sets that can exist in LDPC codes is available}\cite{Vasic}{. This work is general in the sense that we obtain converse results (lower bounds) for a given code object and decoder realization. This can help significantly at the code selection stage of system design.} 


		
	\end{itemize} 
	
	\section{Results and Discussion} \label{Resz}
	
	In this section,  the code-independent approximate lower bounds $\hat\lambda$ and FER performance estimates $N\hat\lambda$ are evaluated for several absorbing sets with SPA and MSA decoders and different quantizers and  compared to simulated FER results for various codes containing those absorbing sets.  In all cases, BPSK-modulated signaling over the AWGN channel is considered. To determine those  quantized channel LLR values that cause decoding failure, which is necessary to compute the bounds, we allowed the absorbing set decoder to perform up to 200 iterations;  however, it was observed that the determination of decoding failure typically did not change after many fewer iterations, usually around 20 in the case of Row Set I. A maximum of 200 iterations was also allowed for the simulated FER code results.

	\subsection{ Approximate Lower Bounds $\hat\lambda$ for Various Absorbing Sets}
\begin{figure}[t]
		\centering
		\includegraphics[width=\columnwidth]{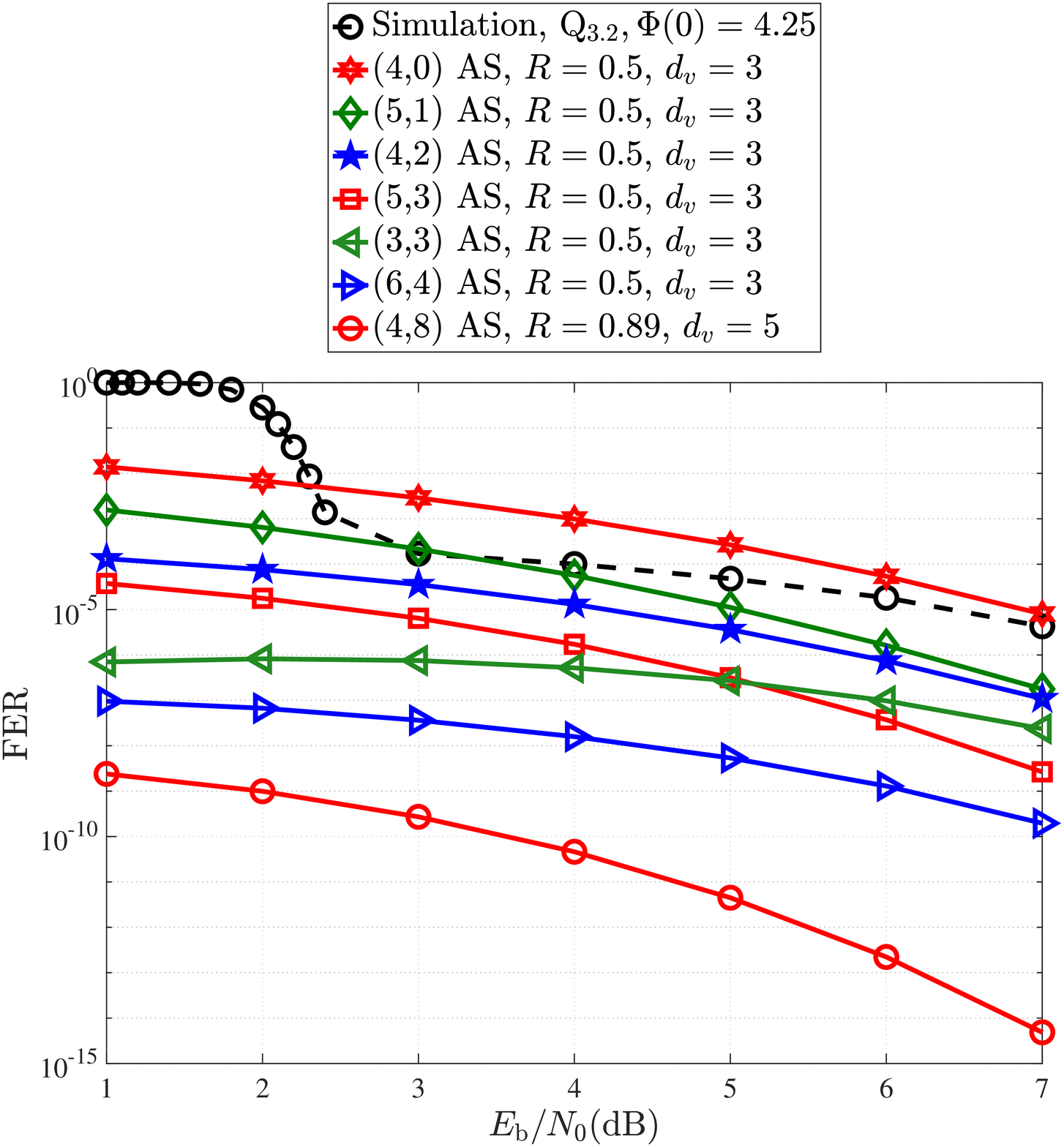}\\
		\caption{Approximate lower bound $\hat{\lambda}$ based on various absorbing sets  with different variable node degrees $d_v$ and different rates $R$. 
			The SPA is used and the quantization scheme is $\text{Q}_{3.2}$ with $\Phi(0)=4.25$.}\label{R1z}	
	\end{figure}

	In this section, the approximate lower bound $\hat\lambda$ for different absorbing sets is evaluated for the SPA decoder with a $\text{Q}_{3.2}$ quantizer and $\Phi(0) = 4.25$, where the code rate is taken into account by proper shifting of the FER vs. ${E_\text{b}}/{N_\text{0}}\text{(dB)}$ curve, as discussed in Section~\ref{dd}.\footnote{It is important to choose the parameter $\Phi(0)$, which is undefined, properly  for the quantized SPA decoder, since its value can affect performance (see~\cite{dol09} for details). }
	Before presenting the results, we note that, for some $(a,b)$ absorbing sets, there may be more than one configuration with these parameters. 
	For example, there exist distinct  $(5,3)$ absorbing sets with girth four, six, or eight (see, \textit{e.g.},~\cite{Vasic}).
	In this case, the absorbing sets with the largest possible girth are chosen for analysis, since good codes are designed to have large girth.
	In Fig. \ref{R1z}, $\hat\lambda$ for the SPA decoder with a $\text{Q}_{3.2}$ quantizer and $\Phi(0) = 4.25$ is depicted for the $(4,0)$, $(5,1)$, $(4,2)$, $(5,3)$, and $(6,4)$ absorbing sets with rate $R = 0.5$ and fixed variable node degree $d_v = 3$, \textit{i.e.}, each variable node in the absorbing set is connected to $3$ check nodes of the absorbing set. 
	The results  demonstrate that increasing $a$ or $b$ leads to a lower $\hat{\lambda}$, as expected. The simulated performance of the randomly constructed $R = 0.5$ $(3,6)$-regular code with length $n = 4000$ from~\cite{MacKay} is also depicted in Fig. \ref{R1z} for the same decoder and quantizer parameters. A $(4,2)$ absorbing set (or LETS), as shown in Fig.~\ref{Aset4_2}, exists in this code, and we see that the calculated $\hat{\lambda}$ provides a lower bound of its performance.
	
	We have also computed $\hat{\lambda}$ for some absorbing sets with fixed variable node degrees $d_v >3$. The results confirm that these absorbing sets typically have a lower $\hat{\lambda}$  than for absorbing sets with $d_v =3$. (Again, this is to be expected, since larger variable node degrees generally correspond to stronger codes,) As an illustration, the bound for the $(4,8)$ absorbing set with $d_v = 5$,  which was identified as the dominant absorbing set for the length $n=2209$ and rate $R = 42/47 = 0.89$ $(5,47)$ array code using the $\text{Q}_{3.2}$ quantizer with $\Phi(0) = 4.25$ in~\cite{dol09}, is seen to be much lower than any of the bounds for the $d_v = 3$ absorbing sets shown in Fig.~\ref{R1z}.\footnote{These results do not yet consider the multiplicity of the absorbing set, which plays a significant role for structured codes.}
	
	\subsection{FER Performance Estimates $N\hat{\lambda}$ for Various Codes}
	In this section, the randomly constructed code of~\cite{MacKay}, several array codes~\cite{fan}, a Tanner code~\cite{Tan04}, and a Euclidean Geometry (EG) code~\cite{Kou01} are considered. Based on the dominant absorbing set  for each code, their estimated FER performance $N\hat{\lambda}$ is evaluated.\footnote{A code might have different dominant absorbing sets, depending on the decoding algorithm. Also, the dominant absorbing set might depend on SNR. Here, we consider the absorbing sets that are dominant in the error floor (high SNR) region. }
	
\begin{figure}[t]
		\centering
		\includegraphics[width=\columnwidth]{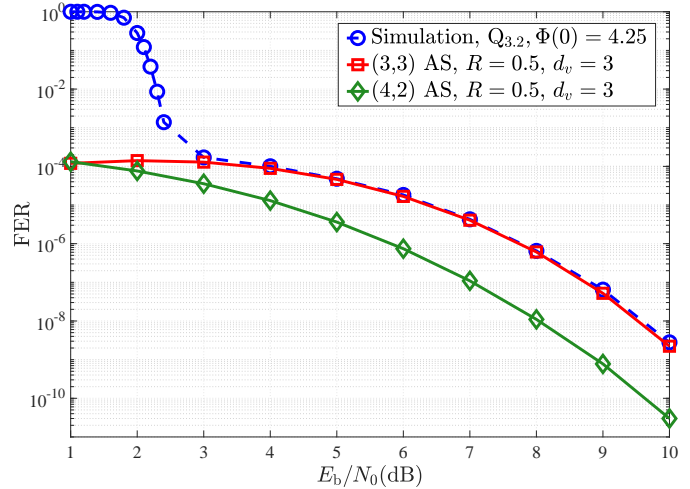}\\
		\caption{Simulation results (dashed blue line) for a $(3,6)$ code with length $n = 4000$ and rate $R = 0.5$ of~\cite{MacKay03} and estimated FER performance $N\hat{\lambda}$ (solid lines) based on a $(4,2)$ absorbing set and a $(3,3)$ absorbing set with multiplicities $N = 1$ and $N = 171$, respectively, decoded using the  SPA with a $\text{Q}_{3.2}$ uniform quantizer. }\label{11az}
	\end{figure}

	We first consider again the $(3,6)$ randomly constructed code of length $n = 4000$ and rate $R = 0.5$~\cite{MacKay}. Fig.~\ref{11az} shows the simulated FER performance obtained with an SPA decoder and a 6-bit uniform quantizer. Also shown are the FER estimates $N\hat{\lambda}$ for the $(4,2)$ and $(3,3)$ absorbing sets, where the multiplicities $N = 1$ and $N = 171$, respectively, were obtained from~\cite{yoones16}. It is observed that, even though a single $(3,3)$ absorbing set (also classified as a LETS) is much less harmful than a single $(4,2)$ absorbing set (see Fig.\ref{R1z}), when the multiplicities are considered the $(3,3)$ absorbing set is dominant in the error floor region for this decoder.\footnote{We see from the results in Fig.~\ref{11az} that it is not unusual for the simulated performance to diverge from the estimated performance in the waterfall region, where the concept of a single dominant absorbing set is no longer relevant.}
	
\begin{figure}[t]
		\centering
		\includegraphics[width=\columnwidth]{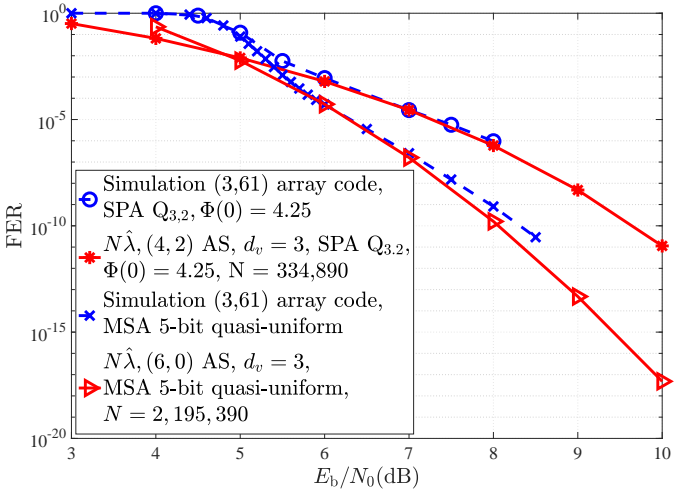}\\
		\caption{ Simulation results (dashed lines)  for the $(3,61)$ array code with length $n = 3721$ and rate $R = 0.9514$ and estimated FER performance $N\hat{\lambda}$ (solid lines) based on a $(4,2)$ absorbing set and a $(6,0)$ absorbing set, respectively, decoded using the SPA with a $\text{Q}_{3.2}$ uniform quantizer and  the MSA with 5-bit quasi-uniform quantizer.}\label{1az}
	\end{figure}

	In Fig.~\ref{1az}, the simulated FER performance (dashed blue circles) is shown for the  $(3,61)$ array code~\cite{fan} of length $n = 3721$ and rate $R = 0.9514$  when decoded using the SPA and a 6-bit uniform quantizer. Also shown is the estimated FER performance $N\hat{\lambda}$ (solid red dots), where $\hat{\lambda}$ was computed for a $(4,2)$ absorbing set and the same decoder with multiplicity $N = 334,890$ obtained from~\cite{dolecek10}. We see that the performance estimate is accurate, since this absorbing set is dominant. The simulated performance is also shown for the MSA decoder and a 5-bit quasi-uniform (dashed blue crosses). We observe here that the MSA outperforms the SPA, as previously noted in~\cite{siegel14}, for this code. The estimated FER performance $N\hat{\lambda}$ is shown (solid red triangles), where $\hat{\lambda}$ was computed for a $(6,0)$ absorbing set and the same MSA decoder, with multiplicity $N = 2,195,390$ obtained from~\cite{Liu10}. Again, the FER estimate closely tracks the simulated performance. 	

\begin{figure}[t]
		\centering
		\includegraphics[width=\columnwidth]{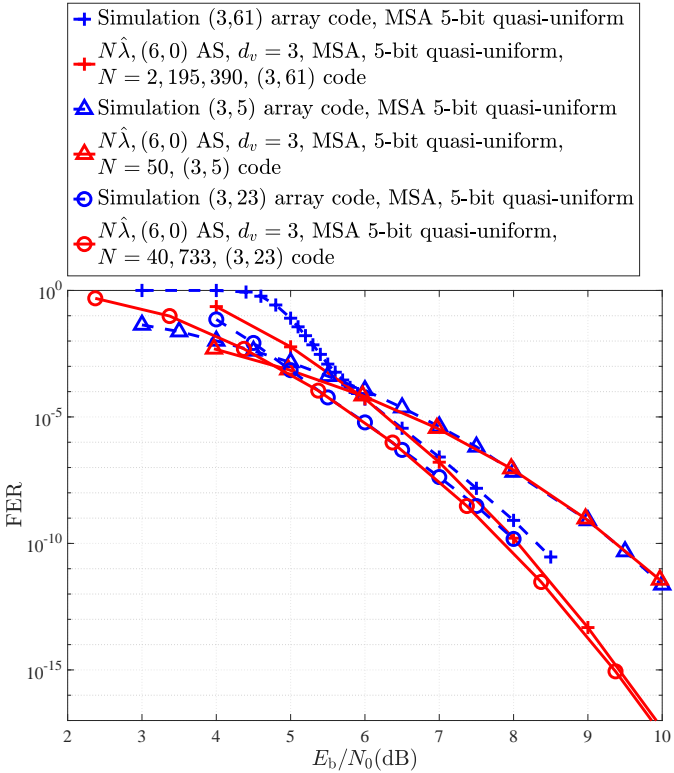}\\
		\caption{Simulated performance versus estimated FER performance based on the dominant $(6,0)$ absorbing set (AS) for the rate $R = 0.48$, $0.8733$, and $0.9514$, length $n = 25$, $529$, and $3721$, $\left(3,5\right)$, $\left(3,23\right)$, and $\left(3,61\right)$ array codes, respectively, with a 5-bit quasi-uniform quantizer and an MSA decoder.}\label{5_5}
	\end{figure}

	In Fig.~\ref{5_5}, three array codes with different parameters are examined. For an MSA decoder with a $5$-bit quasi-uniform quantizer, the (6,0) absorbing set is dominant in all cases. The estimated FER performance is evaluated based on a normalized rate and then shifted as described in Section~\ref{dd}. We again see that  the FER estimates closely track the error floor performance of all three codes. {As a final array code example, we note that the FER estimate of }\eqref{eqn:d8a}{ can be shown to be consistent with the prediction, bounds, and hardware experiments of (5,47)-regular array codes with a $\text{Q}_{3.2}$ quantizer}\cite{dolecek10}{. For this code, the $(4,8)$ absorbing set is known to be dominant. The approximate lower bound $\hat{\lambda}$ for a $(4,8)$ absorbing set and a $\text{Q}_{3.2}$ quantizer is shown in Fig. }\ref{R1z}{, from which the performance estimate can be obtained using the multiplicity $N=304,842$.}\footnote{{The authors would like to thank H. Esfahanizadeh and L. Dolecek for providing the multiplicity of the $(4,8)$ absorbing sets in the $(5,47)$-regular array-based LDPC code.}}
	
\begin{figure}[t]
			\centering
			\includegraphics[width=\columnwidth]{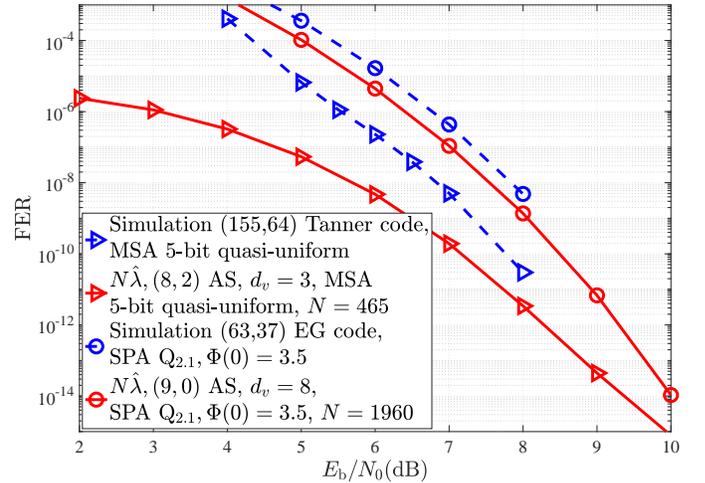}\\
			\caption{Simulated performance of the  $(155,64)$ Tanner code with 5-bit quasi-uniform quantized MSA decoding compared to its estimated performance $N\hat\lambda$  obtained based on the (8,2) absorbing set and of the  (63,37) EG code with 4-bit quantized SPA decoding compared to its  estimated performance $N\hat{\lambda}$  based on the (9,0) absorbing set.}\label{tanner-eg}
		\end{figure}

	Finally, in Fig.~\ref{tanner-eg}, two different codes with different decoding algorithms, quantizers, and dominant absorbing sets are considered. The first one is the rate $R = 0.413$, $(155,64)$ Tanner code~\cite{Tan04}, decoded with the MSA decoder and a 5-bit quasi-uniform quantizer, where the $N =465$ $(8,2)$ absorbing sets (LETS) are dominant. The second one is the rate $R = {0.587}$, $(63,37)$ EG code~\cite{Kou01}, decoded with the SPA decoder and a 4-bit uniform quantizer, where the $N = 1960$ $(9,0)$ absorbing sets (codewords) are dominant. The  estimated FER performance of both codes is depicted along with their simulated performance. We again see that the estimates closely track the simulated FER in the error floor (high SNR) region. 
	For the Tanner code, the $(8,2)$ absorbing set is dominant only at high SNRs, which explains the fact that  the estimate diverges from the simulation at low SNRs. For the EG code, the $(9,0)$ absorbing set is dominant over a wide range of SNRs. Therefore, the estimate closely follows the simulation at all SNRs. 

	\section{Conclusions}\label{Concludez}
	In this paper, we presented approximate lower bounds $\hat{\lambda}$ and  FER performance estimates $N\hat{\lambda}$ for LDPC codes known to contain a given absorbing set for quantized MP decoders and the binary-input AWGN channel. The bounds and estimates are  general, in the sense that they apply to any code containing a given  absorbing set and only depend on the rate of the code. For a given absorbing set, we used the concept of an absorbing set decoder to derive a lower bound on its FER by finding the set of channel input patterns to the absorbing set that are not correctly decoded under any circumstances  imposed by the LLRs coming from outside the absorbing set.  We then showed that, instead of considering all possible realizations of the LLRs coming into the check nodes of the absorbing set, it  suffices to examine only a few  realizations to obtain a good approximate lower bound on the FER, thus making its computation much simpler. We also showed that, if the multiplicity of the dominant absorbing set in a code is known, an accurate estimate of the code's FER performance in the error floor (high SNR) region is obtained. Finally, using various examples, we showed that the approximate lower bound and the FER performance estimates, which can be evaluated much faster than performing conventional Monte-Carlo simulations,
	are useful tools in predicting the high SNR behavior of quantized LDPC decoders.	
	

	%
	%

	\ifCLASSOPTIONcaptionsoff
	\newpage
	\fi
	\bibliographystyle{IEEEtran}

\begin{thebibliography}{10}
\providecommand{\url}[1]{#1}
\csname url@samestyle\endcsname
\providecommand{\newblock}{\relax}
\providecommand{\bibinfo}[2]{#2}
\providecommand{\BIBentrySTDinterwordspacing}{\spaceskip=0pt\relax}
\providecommand{\BIBentryALTinterwordstretchfactor}{4}
\providecommand{\BIBentryALTinterwordspacing}{\spaceskip=\fontdimen2\font plus
\BIBentryALTinterwordstretchfactor\fontdimen3\font minus
  \fontdimen4\font\relax}
\providecommand{\BIBforeignlanguage}[2]{{%
\expandafter\ifx\csname l@#1\endcsname\relax
\typeout{** WARNING: IEEEtran.bst: No hyphenation pattern has been}%
\typeout{** loaded for the language `#1'. Using the pattern for}%
\typeout{** the default language instead.}%
\else
\language=\csname l@#1\endcsname
\fi
#2}}
\providecommand{\BIBdecl}{\relax}
\BIBdecl

\bibitem{gallager1}
R.~G. Gallager, ``Low-density parity-check codes,'' \emph{IRE Trans. Inform.
  Theory}, vol. IT-8, Jan. 1962.

\bibitem{MacKay03}
D.~J.~C. MacKay and M.~S. Postol, ``Weaknesses of {M}argulis and
  {R}amanujan-{M}argulis low-density parity-check codes,'' \emph{Electron.
  Notes Theoretical Comput. Sci.}, vol.~74, no.~8, pp. 97--104, Oct 2003.

\bibitem{richardson03}
T.~J. Richardson, ``Error floors of {{LDPC}} codes,'' \emph{2003 Allerton Conf.
  Communications, Control and Computing}, pp. 1426-- 1435, Nov. 2003.

\bibitem{dolecek10}
L.~Dolecek, Z.~Zhang, V.~Anantharam, M.~Wainwright, and B.~Nikolic, ``Analysis
  of absorbing sets and fully absorbing sets of array-based {{LDPC}} codes,''
  \emph{IEEE Trans. Inf. Theory}, vol.~56, no.~1, pp. 181--201, Jan. 2010.

\bibitem{tomasoni17}
A.~Tomasoni, S.~Bellini, and M.~Ferrari, ``Thresholds of absorbing sets in
  low-density parity-check codes,'' \emph{IEEE Trans. Commun.}, vol.~65, no.~8,
  pp. 3238--3249, Aug 2017.

\bibitem{Kyung12}
G.~B. {Kyung} and C.~{Wang}, ``Finding the exhaustive list of small fully
  absorbing sets and designing the corresponding low error-floor decoder,''
  \emph{IEEE Trans. Commun.}, vol.~60, no.~6, pp. 1487--1498, June 2012.

\bibitem{LETS}
M.~{Karimi} and A.~H. {Banihashemi}, ``On characterization of elementary
  trapping sets of variable-regular {LDPC} codes,'' \emph{IEEE Trans. Inf.
  Theory}, vol.~60, no.~9, pp. 5188--5203, Sep. 2014.

\bibitem{LETS1}
S.~{Laendner}, T.~{Hehn}, O.~{Milenkovic}, and J.~B. {Huber}, ``The trapping
  redundancy of linear block codes,'' \emph{IEEE Trans. Inf. Theory}, vol.~55,
  no.~1, pp. 53--63, Jan 2009.

\bibitem{LETS2}
O.~{Milenkovic}, E.~{Soljanin}, and P.~{Whiting}, ``Asymptotic spectra of
  trapping sets in regular and irregular {LDPC} code ensembles,'' \emph{IEEE
  Trans. Inf. Theory}, vol.~53, no.~1, pp. 39--55, Jan 2007.

\bibitem{LETS3}
Y.~{Zhang} and W.~E. {Ryan}, ``Toward low {LDPC}-code floors: a case study,''
  \emph{IEEE Trans. Commun.}, vol.~57, no.~6, pp. 1566--1573, June 2009.

\bibitem{dolecek09}
L.~Dolecek, P.~Lee, Z.~Zhang, V.~Anantharam, B.~Nikolic, and M.~Wainwright,
  ``Predicting error floors of structured {{LDPC}} codes: deterministic bounds
  and estimates,'' \emph{IEEE J. on Sel. Areas in Commun.}, vol.~27, no.~6, pp.
  908--917, August 2009.

\bibitem{alerton17}
H.~{Hatami}, D.~G.~M. {Mitchell}, D.~J. {Costello}, and T.~{Fuja}, ``Lower
  bounds for quantized {LDPC} min-sum decoders based on absorbing sets,'' in
  \emph{Proc. \nth{55} Annual Allerton Conf Commun., Control, and Computing},
  Oct 2017, pp. 694--699.

\bibitem{bani13}
H.~Xiao, A.~Banihashemi, and M.~Karimi, ``Error rate estimation of low-density
  parity-check codes decoded by quantized soft-decision iterative algorithms,''
  \emph{IEEE Trans. Commun.}, vol.~61, no.~2, pp. 474--484, Feb. 2013.

\bibitem{sun04}
J.~Sun, O.~Y. Takeshita, and M.~P. Fitz, ``Analysis of trapping sets for {LDPC}
  codes using a linear system model,'' in \emph{Proc. \nth{42} Annu. Allerton
  Conf., Monticello, IL, USA,}, Sep./Oct. 2004, pp. 1701--1702.

\bibitem{schelegel13}
S.~Zhang and C.~Schlegel, ``Controlling the error floor in {{LDPC}} decoding,''
  \emph{IEEE Trans. Commun.}, vol.~61, no.~9, pp. 3566--3575, Sept. 2013.

\bibitem{butler14}
B.~K. Butler and P.~H. Siegel, ``Error floor approximation for {LDPC} codes in
  the {AWGN} channel,'' \emph{IEEE Trans. Inf. Theory}, vol.~60, no.~12, pp.
  7416--7441, Dec 2014.

\bibitem{siegel14}
X.~Zhang and P.~Siegel, ``Quantized iterative message passing decoders with low
  error floor for {{LDPC}} codes,'' \emph{IEEE Trans. Commun.}, vol.~62, no.~1,
  pp. 1--14, Jan. 2014.

\bibitem{Liu10}
H.~Liu, L.~Ma, and J.~Chen, ``On the number of minimum stopping sets and
  minimum codewords of array {LDPC} codes,'' \emph{IEEE Commun. Lett.},
  vol.~14, no.~7, pp. 670--672, July 2010.

\bibitem{yoones16}
Y.~Hashemi and A.~H. Banihashemi, ``New characterization and efficient
  exhaustive search algorithm for leafless elementary trapping sets of
  variable-regular {LDPC} codes,'' \emph{IEEE Trans. Inf. Theory}, vol.~62,
  no.~12, pp. 6713--6736, Dec 2016.

\bibitem{dol09}
Z.~Zhang, L.~Dolecek, B.~Nikolic, V.~Anantharam, and M.~Wainwright, ``Design of
  {{LDPC}} decoders for improved low error rate performance: quantization and
  algorithm choices,'' \emph{IEEE Trans. Commun.}, vol.~57, no.~11, pp.
  3258--3268, Nov. 2009.

\bibitem{angarita14}
F.~Angarita, J.~Valls, V.~Almenar, and V.~Torres, ``Reduced-complexity min-sum
  algorithm for decoding {LDPC} codes with low error-floor,'' \emph{IEEE Trans.
  Circuits Syst., I}, vol.~61, no.~7, pp. 2150--2158, July 2014.

\bibitem{fan}
J.~L. Fan, ``Array codes as low-density parity-check codes,'' in \emph{Proc.
  \nth{2} Int. Symp. Turbo Codes}, 2000, pp. 545--546.

\bibitem{isit19}
H.~{Hatami}, D.~G.~M. {Mitchell}, D.~J. {Costello}, and T.~{Fuja}, ``A modified
  min-sum algorithm for quantized {LDPC} decoders,'' in \emph{Proc. IEEE Int.
  Symp. Inf. Theory}, 2019.

\bibitem{Vasic}
\url{http://www2.engr.arizona.edu/~vasiclab/project.php?id=9}.

\bibitem{MacKay}
\url{http://www.inference.phy.cam.ac.uk/mackay/S0.html}.

\bibitem{Tan04}
R.~M. Tanner, D.~Sridhara, A.~Sridharan, T.~E. Fuja, and D.~J. Costello,
  ``{LDPC} block and convolutional codes based on circulant matrices,''
  \emph{IEEE Trans. Inf. Theory}, vol.~50, no.~12, pp. 2966--2984, Dec 2004.

\bibitem{Kou01}
Y.~Kou, S.~Lin, and M.~P.~C. Fossorier, ``Low-density parity-check codes based
  on finite geometries: a rediscovery and new results,'' \emph{IEEE Trans. Inf.
  Theory}, vol.~47, no.~7, pp. 2711--2736, Nov 2001.

\end{thebibliography}

	%
	%
	%
\end{document}